\documentclass[preprintnumbers,article,amsmath,amssymb,floatfix,10pt,prd,onecolumn,
superscriptaddress,nofootinbib]{revtex4}
\usepackage{bbm}
\usepackage{amsfonts}
\usepackage{mathrsfs}
\usepackage{latexsym}
\usepackage[cp1251]{inputenc}
\usepackage{epsfig}
\usepackage{epstopdf}
\usepackage{graphicx}
\usepackage{amssymb}
\usepackage{amsmath}
\usepackage{dcolumn}
\usepackage{bm}
\usepackage{color}
\usepackage{comment}
\usepackage{xcolor}

\begin{document}
\title{\bf Embedded Class-I Solution of Compact Stars in $f(R)$ Gravity with\\ Karmarkar Condition}
\author{Tayyaba Naz}
\email{tayyaba.naz@nu.edu.pk}\affiliation{National University of Computer and
Emerging Sciences,\\ Lahore Campus, Pakistan.}
\author{Ammara Usman}
\email{ammarausman88@ymail.com}\affiliation{National University of Computer and
Emerging Sciences,\\ Lahore Campus, Pakistan.}
\author{M. Farasat Shamir}
\email{farasat.shamir@nu.edu.pk}\affiliation{National University of Computer and
Emerging Sciences,\\ Lahore Campus, Pakistan.}

\begin{abstract}
This paper's main aim is to investigate the existence of a new classification of embedded class-I solutions of compact stars, by using Karmarkar condition in $f(R)$ gravity background. To achieve that goal, we consider two different models of the $f(R)$ theory of gravity for the static spherically symmetric spacetime by considering anisotropic matter distribution. Further, we employ Karmarkar condition to relate the two components of metric potentials $g_{rr}$ and $g_{tt}$. We assume a particular model for one metric potential and obtain the second one by Karmarkar condition. Moreover, we also calculate the values of constant parameters by using the observational data of these compact stars, namely, $Vela~ X-1,$ $PSR~J1614-2230,$ $4U~1608-52,$ $Cen~X-3$ and $4U~1820-30.$ We perform different physical tests like variational behavior of energy density and pressure components, stability and equilibrium conditions, energy constraints, mass function and adiabatic index to check the viability of $f(R)$ gravity models. All these physical attributes indicate the consistent behavior of our models. Our investigation also suggests that $f(R)$ theory of gravity appears as a suitable theory in describing the viability of a new classification of embedded class-I solutions of compact objects.
\end{abstract}

\maketitle

\section{Introduction}
Compact objects are usually designated as neutron stars, white dwarfs and black holes in astrophysics. Neutron stars and white dwarfs are born in the result of gravitational collapse which generally happen due to degeneracy pressure of the relativistic objects. These object emerged huge densities values because these stars were considered to be massive objects but volumetrically smaller. Although, we do not know the precise characteristics of these kinds of compact stars, but these objects are presumed to be heavy stars having tiny radius. Every kinds of stellar objects are mostly acknowledged as degenerate stars, except black holes. In general relativity ($GR$) the study of configured equilibrium of compact stars is valuable because of the massive nature and high densities of these objects. The study of compact objects in the framework of $GR$ and modified gravitational theories has always been considered as a topic of great interest in astrophysics. To investigate celestial compact structure modeling, we need an exact solution to Einstein field equations (EFE). In $1916,$ Schwarzschild \cite{Sch} obtained the EFE solution for the interior structure of compact objects. In this regard, various modified gravitational theories have been presented in order to obtain the complicated exact solutions of EFE. In the framework of observational data, Tolman \cite{Tolman} and Oppenheimer \cite{Oppen} investigated some realistic models of non-transversable stellar objects and claimed that the physical characteristics of these objects represent the relationship between the internal pressure and the gravitational force which eventually leads in a state of equilibrium structures. In the study of stars internal configuration, this phenomenon has considerable importance and provides realistic results in various occasions. Further, compact celestial structure was examined by Baade and Zwicky \cite{Baa}. According to their study the supernova might transform into smaller compact objects after the observation of strongly magnetized spinning neutrons. Moreover, Ruderman \cite{Rud} identifies for the first time that at the center of the stellar object the nuclear density exhibits anisotropic behavior.\\
A perfect fluid was assumed to be a source of celestial structures for the creation of stellar objects. These objects are commonly defined as ultra-dense, isotropic, celestial bodies that seem to be spherically symmetric. Whereas, isotropy is assumed as a desired attribute, still it does not have a general feature of compact stars. The concept of non-zero anisotropy in a stellar configuration was first presented by Bowers and Liang \cite{Bow}. Moreover, Ruderman \cite{Rud} investigated the possibility that stars might come with huge densities $10^{15}~g/cm^{3},$ when the nuclear matter is anisotropic in nature. In fact, anisotropic matter distribution plays a crucial role in narrating the inner representation and development phase of the relativistic stellar objects. In anisotropic fluid distribution background, a lot of literature is available \cite{Bhar1,Bhar2,Mau,MH,Her1}. In anisotropic distribution, the pressure component of the fluid sphere divides into two components, radial and transverse.\\
One of the most promising topics of modern research in cosmology is the accelerated expansion of the universe. Cosmologists argue that the accelerating expansion of the universe depends on dark energy and dark matter which retains negative pressure. As a substitute of $GR$, different gravitational theories have been presented to unfold the mystery behind dark energy issues. These gravitational theories are recognized as modified theories of gravity. Some of these modified theories are $f(R),~f(R,T),~f(G),~f(R,G).$ Among these valuable theories, $f(R)$ is one of the most simplest and popular theory, obtained as an arbitrary function of Ricci scalar. This theory was proposed by Buchdahl \cite{Buch}. Later on,  Nojiri and Odintsov \cite{Odintsov1} demonstrated some models of $f(R)$ theory of gravity by placing curvature as a function of Ricci scalar and the outcomes of their considered models are quite viable and stable. Further, Starobinsky \cite{Star} presented an interesting class of $f(R)$ theory of gravity models that showed the physically acceptable results in laboratory testing of the solar structure. Some $f(R)$ gravity model was proposed by Hu and Sawicki \cite{Hu} by ignoring the cosmological constant and their study evident some interesting results in regard to accelerating expansion phenomena. The viability of physical attributes of compact stars by taking exponential type models of $f(R)$ theory of gravity was discussed by Cognola et al. \cite{Cog}. In modified theories of gravity beyond $GR$ and its Hilbert-Einstein action, diffeomorphism invariance and the Bianchi identities violation has attracted a lot of interest. Hamity and Barraco \cite{Ha} derived the generalized Bianchi identities for the non-linear $f(R)$ gravity to throw some light on the issue that $f(R)$ theory of gravity generates higher than second order equations of motion and violates Bianchi identities. Further, Wang et al. \cite{Wa} confirmed the local energy-momentum conservation of Bianchi identities by establishing the equivalence relation between Palatini $f(R)$ and the Brans-Dicke gravity. Moreover, Koivisto \cite{Ko} explored a composition of $f(R)$ gravity and the generalized Brans-Dicke gravity and claimed covariant conservation from both the metric tensors and the Palatini variational techniques.\\
For physically stable models, one could use an analytical approach of EFE and assume the family of a four dimensional manifold and transform it into Euclidean space. The embedding family of curved geometry into geometries of higher dimensions is considered to create various new exact solutions in astrophysical stellar systems. Schlai \cite{S1} designated the embedding issue on geometrically important spacetimes for the very first time. In this regard, Nash \cite{Nash} presented the isometric embedding theorem. Several features of anisotropic compact objects by employing embedding class one approach have been discussed in literature \cite{101,103,104,105,106,107,108,109}. A class of non-static fluid distribution along with non-vanishing acceleration was studied by Gupta and Gupta \cite{GG}. Further, Gupta and Sharma \cite{GS} assumed plane symmetric metric to explore the embedding class-I solutions of non-static perfect fluid. The embedding class constraint yields a differential equation in static spherically symmetric geometry relating the two components of metric potentials, known as Karmarkar condition \cite{Kar}. A lot of work \cite{Mau2,Abb,Wah,Mus} has been done related to Karmarkar condition, defined as $R_{1414}R_{2323}=R_{1212}R_{3434}+R_{1224}R_{1334}$. The Karmarkar condition develops a connection between two parts of metric tensors $g_{rr}$ and $g_{tt}$ for a spherical static symmetric fluid distribution. Maurya et al. \cite{Mau3} and Bhar et al. \cite{Bhar} studied the EFE by adopting the Karmarkar condition and composing numerous  classes of embedded class-I solutions. They also observed that these outcomes exhibit stable nature and might be helpful in exploring the internal structure of the stellar objects. Later, in the background of $f(G)$ gravity, Sharif and Saba \cite{Saba} investigated the charged anisotropic solutions  and their derived solutions are physically consistent and stable. In this regard, a class of embedded solutions by adopting Karmarkar condition is recently presented by Upreti et al. \cite{Up}.\\
In the background of $GR$, Maurya et al. \cite{102} probe an anisotropic compact star using embedding class-I approach. They studied different features of compact stars in the presence of anisotropic fluid distribution and claimed that obtained outcomes describe the internal core of stellar objects. Further, Bhar et al. \cite{Bhar4} presented a new relativistic anisotropic compact star model which is physically acceptable for embedding spacetime and can be used to describe the interior solution of stellar objects. Moreover, the relativistic model for anisotropic compact stars in $GR$ background was studied by Prasad et al. \cite{Pra}. Their results indicate that by adopting embedding class-I condition the obtained relativistic stellar structure is physically reasonable. Recently, Mustafa et al. \cite{Sha2} introduced new exact solutions of EFE in the framework of Bardeen black hole spacetime by using the well known Karmarkar condition. Their chosen model demonstrated the well-behaved nature under the particular values of the parameters. By getting motivated from literature, in this particular paper, we extended the concept of Bhar et al. \cite{Bhar4} in the background of $f(R)$ theory of gravity and analyzed the physical attributes of the compact stars, namely, $Vela~ X-1,$ $PSR~J1614-2230,$ $4U~1608-52,$ $Cen~X-3$ and $4U~1820-30.$ To meet our goal, we assume two different $f(R)$ theory of gravity models by employing the Karmarkar condition in the framework of anisotropic pressure. We consider a particular model for one of the metric potential $g_{rr}$ and by adopting the Karmarkar condition, we construct the second metric potential $g_{tt}$ of spherically symmetric spacetime.\\
The outline of the current article is organized as follows: In the next segment, the $f(R)$ field equations have been developed by using Karmarkar condition with an anisotropic matter source. Two of viable $f(R)$ gravity models along with boundary constraints have been introduced in segment III. In section IV, we compute the constant values by using the matching conditions. In section V, we discuss the physical attributes of the compact stars in detail. In the last portion, we provide the final verdict and conclusion of our study.
\section{Modified Field Equations}
First, we are proceeding to develop the field equations in $f(R)$ gravity context. For this purpose, we consider the action of $f(R)$ gravity \cite{Nod} defined as
\begin{equation}\label{0}
  S= \int\Big[\frac{f(R)}{2\kappa} + \mathcal{L}_{m}\Big]\sqrt{-g}d^4x.
\end{equation}
Here, $f$ is a function of Ricci scalar and $\mathcal{L}_{m}$ is the Lagrangian matter. Varying the action identified in (\ref{0}) regarding the metric potential $g_{\eta \zeta},$ we obtain the succeeding $f(R)$ gravity field equation
\begin{equation}\label{1}
FR_{\eta \zeta}- \frac{1}{2}f(R)g_{\eta \zeta}-\nabla_{\eta}\nabla_{ \zeta}F+g_{\eta \zeta}\Box F = \kappa T_{\eta \zeta}.
\end{equation}
Thus, $F= \frac{df(R)}{dR}.$ Although, $\nabla_{\eta}$ and $\Box$ indicate covariant derivative and D'Alembertian notation, i.e. $\Box\equiv\nabla_{\eta}\nabla^{\zeta}$.
The stress-energy momentum tensor is presented as $$T_{\eta \zeta}=(\rho+p_{r})u_{\eta}u_{ \zeta}+p_{t}g_{\eta \zeta}+(p_{r}-p_{t})\mathcal{X}_{\eta}\mathcal{X}_{ \zeta},$$
where $\rho$, $p_{r}$ and $p_{t}$ represent the energy density, radial and transverse pressure, respectively. Here, $u_{\eta}$ and $\mathcal{X}_{\eta}$ are four velocity vectors, which satisfy the relations $u^{\eta}u_{\eta}=-\mathcal{X}^{\eta}\mathcal{X}_{\eta}=1$. Further, we assume a static spherically symmetric line element defined by
\begin{equation}\label{2}
ds^{2}=e^{\nu(r)}{dt}^{2}-e^{\lambda(r)}{dr}^{2}-r^{2}({d\theta}^{2}+sin^{2}\theta{d\phi}^{2}).
\end{equation}
Here, ${\nu(r)}$ and ${\lambda(r)}$ are the functions of radial coordinate only. We assume that our stress-energy tensor is anisotropic in nature and relativistic geometrized units $G=c=1$, which implies that $\kappa=\frac{8\pi G}{c^{4}}$ reduces to $8\pi.$ The $f(R)$ gravity field equations for spacetime (\ref{2}) yield
\begin{equation}\label{3}
8\pi\rho= e^{-\lambda}F\big(\frac{\nu^{''}}{2}+\frac{\nu^{'}}{r}+\frac{\nu^{'}(\nu^{'}-\lambda^{'})}{4}\big)- \frac{1}{2}f(R)-e^{-\lambda}F^{''}-e^{-\lambda}F^{'}\big(\frac{\lambda^{'}}{2}+\frac{2}{r}\big),
\end{equation}
\begin{equation}\label{4}
8\pi p_{r}= e^{-\lambda}F\big(-\frac{\nu^{''}}{2}+\frac{\lambda^{'}}{r}-\frac{\nu^{'}(\nu^{'}-\lambda^{'})}{4}\big)+ \frac{1}{2}f(R)+e^{-\lambda}F^{'}\big(\frac{\nu^{'}+2\lambda^{'}}{2}+\frac{2}{r}\big),
\end{equation}
\begin{equation}\label{5}
8\pi p_{t}= F\big(\frac{1-e^{-\lambda}}{r^{2}}+e^{-\lambda}\frac{(\lambda^{'}-\nu^{'})}{2r}\big)+ \frac{1}{2}f(R)+e^{-\lambda}F^{''}+e^{-\lambda}F^{'}\big(\frac{\nu^{'}+\lambda^{'}}{2}+\frac{1}{r}\big),
\end{equation}
here `prime' symbolized as a $r$ derivative. Whereas, the anisotropic parameter is presented as
\begin{equation}\label{6}
   \Delta = 8\pi p_{t}-8\pi p_{r}. \\
\end{equation}
By utilizing Eqs. (\ref{4}) and (\ref{5}), we receive
\begin{eqnarray}\label{7}
   \Delta&=&  F\big[\frac{1-e^{-\lambda}}{r^{2}}+e^{-\lambda}\big(\frac{\nu^{''}}{2}-\frac{(\nu^{'}+\lambda^{'})}{2r}+\frac{\nu^{'}(\nu^{'}+\lambda^{'})}{4}\big)\big]
+e^{-\lambda}F^{''}-e^{-\lambda}F^{'}\big(\frac{\lambda^{'}}{2}+\frac{1}{r}\big).
\end{eqnarray}
Moreover, the nature of $\Delta$ is attractive if $p_{r}<p_{t}$ and repulsive if $p_{r}>p_{t}$ \cite{A2}. Further, the metric tensor (\ref{2}) represents the embedded class-I family, if it satisfies the Karmarkar condition, i.e.
\begin{equation}\label{8}
 R_{1414}R_{2323}=R_{1212}R_{3434}+R_{1224}R_{1334},
\end{equation}
with $R_{2323}\neq 0$. By employing the well-known Karmarkar condition, the succeeding differential equation derived for metric (\ref{2})
\begin{equation}\label{9}
 \lambda^{'}\nu^{'}-2\nu^{''}-{\nu^{'}}^{2}=\frac{\lambda^{'}\nu^{'}}{1-e^{\lambda}} ,
\end{equation}
where $e^{\lambda}\neq1$. Integrating Eq. (\ref{9}), we get
\begin{equation}\label{10}
  e^{\nu}=\Big[{{\big(A+B\int{\sqrt{e^{\lambda}-1} dr}\big)}^{2}}\Big].
\end{equation}
Here $A$ and $B$ are constant of integration. In order to obtain the embedded class-I solution, we consider a specific model for the component $g_{rr}=e^{\lambda(r)}$ \cite{Bhar4} which is given as
\begin{equation}\label{11}
  e^{\lambda}=1+ \frac{a^{2}r^{2}}{(1+br^{2})^{4}}.
\end{equation}
To demonstrate embedded class-I spacetime solution $a$ and $b$ should be non-zero. Singh and Pant \cite{SP} explored a similar compact star model, in the context of $GR$, by considering the metric potential $e^{\lambda}=1+a^{2}r^{2}(1+br^{2})^{x}$ of embedding class-I solution. In their research, they only assume positive values of $x$ and presented a family of new exact solutions for relativistic anisotropic stellar objects. Later, Bhar et al. \cite{Bhar4} studied this metric potential by taking $x=-4$ and the obtained results are well behaved in all respects in the background of $GR$. Motivated from the work of Bhar et al. \cite{Bhar4}, here we extended the analysis in $f(R)$ gravity context. Manipulating Eqs. (\ref{10}) and (\ref{11}), the metric potential expression becomes
\begin{equation}\label{12}
  e^{\nu}=\Big[{\Big(A-\frac{aB}{2b(1+br^{2})}\Big)}^{2}\Big].
\end{equation}
\section{Viable $f(R)$ Gravity Models}
Here, we consider the two realistic and simple viable $f(R)$ gravity models to analysis the modified field equations presented in Eqs. $(\ref{3})-(\ref{5}).$
\subsection{Model 1}
To study the existence of the compact stellar objects, we first assume an exponential type $f(R)$ gravity model \cite{Cog}
\begin{equation}\label{13}
  f(R)=R+\alpha(e^{-\beta R}-1),
\end{equation}
where, $\alpha$ and $\beta$ are any arbitrary constants. The exponential based $f(R)$ gravity models plays a very crucial role in the context of describing exponential expansion. A class of exponential, realistic modified gravities was presented by Cognola et al. \cite{Cog} in which they argued that this model (\ref{13}) passes all the local tests including non-violation of Newton's law and stability of spherical body solution. Nojiri and Odintsov \cite{Odintsov1} investigated this model (\ref{13}) to describe the early-time inflation and late-time cosmic acceleration in a natural, unified way. It was shown that exponential type models present a realistic dark energy epoch that is compatible with local and observational tests \cite{Lin}. In order to explore the compact stars existence for our proposed model (\ref{13}), the values of the constant parameters are chosen in kind that $\rho$, $p_{r}$ and $p_{t}$ exhibit positive behavior. The physical features of the compact stars for the model (\ref{13}) by using Eqs. $(\ref{3})-(\ref{5})$ can be defined by the following relation as
\begin{eqnarray}
\nonumber &&
\rho= \frac{1}{16 \pi  r h_{1}^2 h_{2}}\Big(-\alpha  (\beta h_{1})^2 \left(r \lambda '+4\right)h_{2} h_{3}h_{5}+2 \alpha r (\beta h_{1})^2 h_{2} h_{3} \left(\beta  h_{5}^2-h_{6}\right)-r h_{1}^2h_{2} \left(\alpha  \left(e^{-\beta  h_{4}}-1\right)+h_{4}\right) ~~~
 \\&&+2 a b B r  h_{3} \left(r h_{1} \lambda '+2 b r^2-6\right) \left(e^{\beta h_{4}}-\alpha \beta \right)\Big),
\end{eqnarray}
\begin{eqnarray}
\nonumber &&
p_{r}= \frac{1}{8 \pi }\Big(\frac{\left(e^{\beta  h_{4}}-\alpha  \beta \right) h_{3} \left(h_{1} \lambda ' \left(2 A b \left(b r^2+1\right)^2-a B\right)+2 a b B r \left(3 b r^2-1\right)\right)}{-r h_{1}^2 h_{2}}+\alpha  \beta ^2 h_{3} h_{5} (\frac{2 a b B r}{-h_{1}h_{2}}+\lambda '~~~
\\&&+\frac{2}{r})+\frac{1}{2} \left(\alpha  \left(e^{-\beta  h_{4}}-1\right)+h_{4}\right)\Big),
\end{eqnarray}
\begin{eqnarray}
\nonumber &&
p_{t}= \frac{1}{16 \pi }\Big(2 \alpha  \beta ^2 h_{3} h_{5} \left(\frac{2 a b B r}{-h_{1}h_{2}}+\frac{\lambda '}{2}+\frac{1}{r}\right)+\frac{\left(1-\alpha  \beta  e^{-\beta  h_{4}}\right) \left(e^{-\lambda} \left(\frac{4 a b B r^2}{h_{1}h_{2}}-2\right)+r e^{-\lambda } \lambda '+2\right)}{r^2}-2 \alpha  \beta ^2 h_{3}~~~
\\&& \left(\beta  h_{5}^2-h_{6}\right)+\alpha  \left(e^{-\beta  h_{4}}-1\right)+h_{4}\Big).
\end{eqnarray}
\subsection{Model 2}
We also discuss the graphical response of the considered stars for the model \cite{Faul}
\begin{equation}\label{14}
  f(R)=R-(1-n)\mu^{2}\Big(\frac{R}{\mu^{2}}\Big)^{n}.
\end{equation}
Here, $\mu$ and $n$ are arbitrary constant parameters. In literature, the model (\ref{14}) used to investigate the effects of cosmic acceleration and the Chameleon mechanism. Faulkner et al. \cite{Faul} established the relation between the scalar tensor gravity and modified $f(R)$ gravity by using different models. According to them, the fundamental feature of this model is conformal coupling. Furthermore, Amendola et al. \cite{Ame} discussed the standard matter dominated epoch for the $f(R)$ dark energy models. They argued that the model (\ref{14}) showed interesting results in supernova tests and revealed well-behaved consistent and stable reactions in the framework of the solar system. For $f(R)$ gravity model under investigation, we assume the appropriate value of constants to obtained the physically acceptable results. By making use of Eq. (\ref{14}), the explicit relation for the energy density, radial and transverse pressures have been found to be:
\begin{eqnarray}
\nonumber &&
\rho= \frac{1}{32 \pi  r h_{4}^3}\Big(-2 h_{4}h_{5}h_{7} h_{8} \left(r \lambda '+4\right)-4 rh_{7} h_{8} \left(h_{4} h_{6}+(n -2) h_{5}^2\right)+ h_{4}^2 \left(h_{7} h_{8}+e^{-\lambda }h_{4}\right) (\left(4-r \lambda '\right) ~~~
\\&&\nu '+2 r \nu ''+r \nu '^2)-2 r h_{4}^3 \left(\mu ^2 (n -1)h_{8} +h_{4}\right)\Big),
\end{eqnarray}
\begin{eqnarray}
\nonumber &&
p_{r}=\frac{1}{32 \pi  r h_{4}^2}\Big(2 h_{5} h_{7}h_{8} \left(2 r \lambda '+r \nu '+4\right) + h_{4} \left(h_{7} h_{8}+e^{-\lambda }h_{4}\right) \left(\lambda ' \left(r \nu '+4\right)-r \left(2 \nu ''+\nu '^2\right)\right)+2 r h_{4}^2 ~~~
\\&&\left(\mu ^2 (n -1)h_{8}+h_{4}\right)\Big),
\end{eqnarray}
\begin{eqnarray}
\nonumber &&
p_{t}=\frac{1}{16 \pi }\Big(\left(r \lambda '+2 e^{\lambda}-r \nu '-2\right)\left(\frac{h_{7} h_{8}}{r^2 h_{4}(n -1)} +\frac{e^{\lambda}}{r^2}\right)+\frac{h_{5} h_{7} h_{8} \left(r \lambda '+r \nu '+2\right)}{r h_{4}^2}+\mu ^2 (n -1) h_{8}+h_{4}~~~
\\&&+\frac{2 h_{7} h_{8} \left(h_{4} h_{6}+(n -2) h_{5}^2\right)}{h_{4}^3}\Big),
\end{eqnarray}
where, $$  h_{1} = (br^{2}+1),~~~h_{2} = aB-2Ab(br^{2}+1),~~~h_{3}= e^{-\lambda-\beta h_{4}},$$
\begin{eqnarray*}
   \nonumber &&h_{4} = \frac{1}{-\left(a^2 r^2+h_{1}^4\right)^2 h_{2}} \Big(2 a (a^4 B r^2-2 a^3 A b r^2+h_{1}-a^2 B \left(b r^2-3\right)~~~
   \\&& +h_{1}^3+2 a A b \left(5 b r^2-3\right)+h_{1}^4-2 b B \left(b r^2-3\right) +h_{1}^6)\Big),
  \end{eqnarray*}
 $$ h_{5} = \frac{\partial h_{4}}{\partial r},~~~h_{6} = \frac{{\partial h_{4}}^{2}}{{\partial}^{2} r},~~~h_{7} = {{\mu}^{2}}n(n-1)^{2}e^{-\lambda},~~~h_{8} = \left(\frac{h_{4}}{\mu^{2}}\right)^{n}.$$
\subsection{Boundary Conditions}
\begin{figure}[h!]
\begin{tabular}{cccc}
\epsfig{file=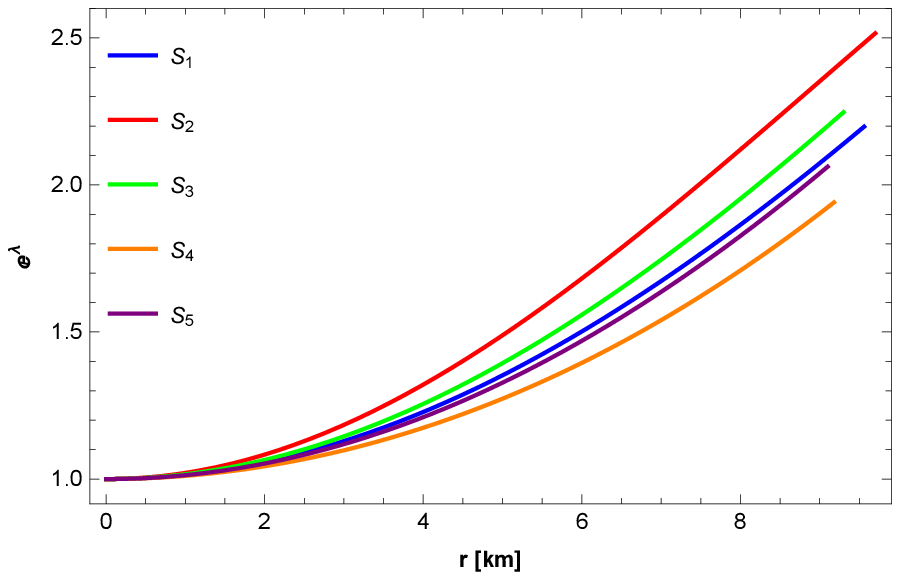,width=0.38\linewidth} &
\epsfig{file=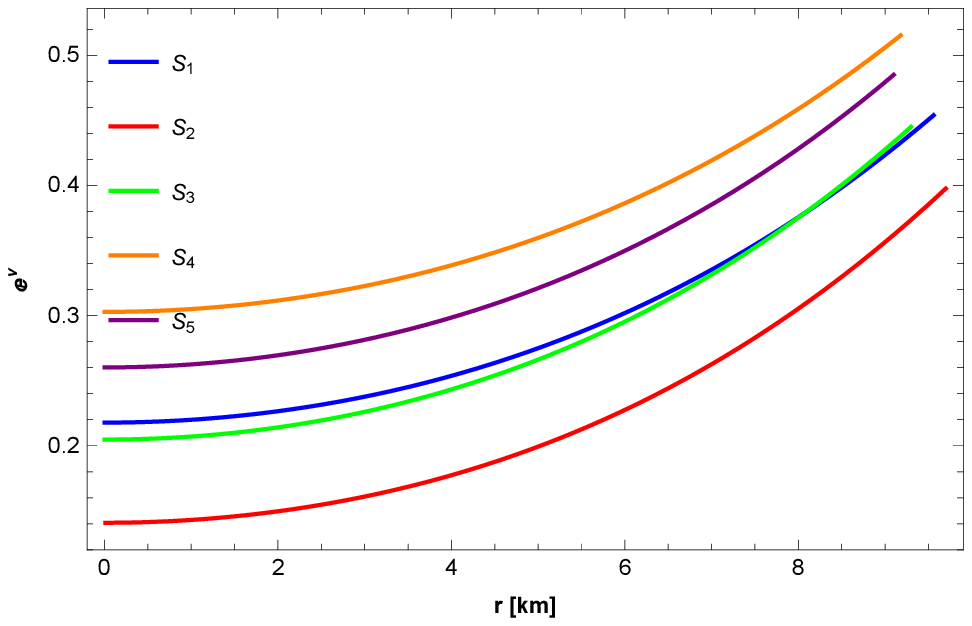,width=0.38\linewidth} &
\end{tabular}
\caption{{Evolution of metric potential $g_{rr}$ and $g_{tt}$}.}
\label{Fig:1}
\end{figure}

Next, we analyze the mandatory conditions for the metric potential, i.e. $e^{\lambda(0)}=1$ and $((e^{\lambda(r)})')_{r=0}=0$. For a well-behaved $f(R)$ theory of gravity model, the graphical response of $g_{rr}$ should not have any type of singularity and the curvature must be regular. From Fig. $\ref{Fig:1},$ one can notice that expression $e^{\lambda(r)}$, given in (\ref{11}), is physically acceptable as both the metric potential shown monotonically increasing behavior, and attains maximum value in the boundary.
Further, we list all the necessary constraints, which must be fulfil for the well-behave nature of the stars.
\begin{itemize}
  \item The $g_{rr}$ and $g_{tt}$ should be singularities free. Also the anisotropic fluid sphere should satisfy $e^{\lambda(r=0)}=1$ and $e^{\nu(r=0)}=constant$.
  \item $\rho,$ $p_{r}$ and $p_{t}$ should always be finite and positive. Further, these illustration must be maximum at center and decreasing toward the boundary.
  \item The gradient of $\rho,$ $p_{r}$ and $p_{t}$ must exhibit negative nature.
  \item The equation of state should obey the necessary condition, i.e. $0<w_{r},~w_{t}<1.$
  \item The velocity sound must lies within $[0,1]$, i.e. $0<{v_{r}}^{2},~{v_{t}}^{2}<1.$
  \item Adiabatic index for anisotropic fluid sphere should be greater the $4/3.$
\end{itemize}
\section{Matching Conditions}
For the stellar compact objects, the intrinsic boundary metric irrespective of the geometry (interior or exterior) will remain the same. This phenomena verify that, whatever the coordinates system covering the surface of the boundary, the components of the metric tensor will remain continuous. For stellar compact objects, the Schwarzschild solution is observed as the most suitable possibility to choose from the various choices of the matching conditions in the context of $GR$. Jebsen-Birkhoff's theorem also ensure that the solution of field equations should be asymptotically flat and remain static for spherical symmetric spacetime. In modified theories of gravity case, the outside solution of star may differ from Schwarzschild solution, when we come to modified TOV equations  \cite{Tolman,Oppen} with zero energy density and pressure. However, this concern may be overcome in modified $f(R)$ gravity with a suitable possibility of $f(R)$ gravity models for non-zero energy density and pressure in Schwarzschild's solution. Due to this phenomena, the Birkhoff theorem fails in modified gravity \cite{Far}. Many research articles have been published on matching conditions \cite{Ast,Cooney,Gan,Mom} and provide some fascinating results for Schwarzschild solution. To compute the field equations solution under the limited boundary conditions at $r=R$, the radial pressure $p_{r}(R)=0$. Here, we match the obtained solution of interior geometry (\ref{2}) with the Schwarzschild exterior geometry, given by
\begin{equation}\label{15}
ds^{2}=\big(1-\frac{2M}{r}\big){dt}^{2}-\big(1-\frac{2M}{r}\big)^{-1}{dr}^{2}-r^{2}({d\theta}^{2}+sin^{2}\theta{d\phi}^{2}),
\end{equation}
The continuity condition for Eqs. (\ref{2}) and (\ref{15}) at the boundary $r=R$ yield
\begin{equation}\label{16}
  {g_{rr}}^{+}={g_{rr}}^{-},~~~~~~~~~~{g_{tt}}^{+}={g_{tt}}^{-},~~~~~~~~~~\frac{\partial{g_{tt}}^{+}}{\partial r}=\frac{\partial{g_{tt}}^{-}}{\partial r}~,
\end{equation}
where ``+" identify the exterior geometry and ``-" identify the interior geometry. With the help of Eqs. (\ref{2}), (\ref{15}) and (\ref{16}), we computed the values of the parameters a, b, A and B, stated as
\begin{equation}\label{17a}
a= \frac{(1+bR^{2})^{2}}{R}\sqrt{\frac{\frac{2M}{R}}{1-\frac{2M}{R}}},~~~~~~~~~~~~~~~~~~~~~~~b= \frac{4M-R}{R^{2}(9R-20M)} ,
\end{equation}
\begin{equation}\label{17b}
A= \frac{aB}{2b(1+bR^{2})}+\sqrt{1-\frac{2M}{R}},~~~~~~~~~~~~~~~~~~ B= (\frac{1}{2R})\sqrt{\frac{2M}{R}}.
\end{equation}
The values of the parameters by assuming the radius and mass of considered compact stars $Vela~ X-1,$ $PSR~J1614-2230,$ $4U~1608-52,$ $Cen~X-3,$ $4U~1820-30,$ are given in Tab. \ref{tab1}.
\begin{table}[ht]
\centering
\caption{Constants values of $a$, $b$, $A$ and $B$ for considered compact stars}.
\begin{tabular}{|p{2.2cm}|p{2.8cm}| p{2.9cm}| p{2.8cm}| p{2.8cm}| p{2.8cm}|}
\hline
\hline
~
\begin{center}
\textbf{Star Model}
\end{center}
    & ~~~ \begin{center}
    $\textbf{Vela~ X-1}$ \\
     $(S_{1})$
    \end{center} & ~~~  \begin{center}
    $\textbf{PSR~J1614-2230}$ \\
     $(S_{2})$
    \end{center} & ~~~ \begin{center}
    $\textbf{4U~1608-52}$ \\
     $(S_{3})$
    \end{center} & ~~~ \begin{center}
    $\textbf{Cen~X-3}$ \\
     $(S_{4})$
    \end{center} & ~~~ \begin{center}
    $\textbf{4U~1820-30}$ \\
     $(S_{5})$
    \end{center}\\
\hline
~~~~ $M~(M_{\Theta})$&~~~ 1.77 $\pm$ 0.08 \cite{Gang}&~~~  1.97 $\pm$ 0.04 \cite{Gang} & ~~~  1.74 $\pm$ 0.01 \cite{Gang}  &~~~1.49 $\pm$ 0.08 \cite{Gang}  &~~~1.58 $\pm$ 0.06 \cite{Gang}    \\
\hline
~~~~ $R~(km)$       &~~~~ 9.56 $\pm$ 0.08       &~~~~ 9.69 $\pm$ 0.2   &~~~~ 9.3 $\pm$ 0.10     &~~~~ 9.178 $\pm$ 0.13   &~~~~ 10.56 $\pm$0.10  \\
\hline
~~~~ $M/R$ &~~~~~ 0.272756        &~~~~~ 0.301209                      &~~~~~  0.277485        &~~~~~ 0.242338         &~~~~~ 0.257507 \\
\hline
~~~~ $a~(km)$      &~~~~~ 0.120435  &~~~~~ 0.144972      &~~~~~ 0.127851        &~~~~~ 0.104090      &~~~~~ 0.115014  \\
\hline
~~~~ $b~(km)$      &~~~~~ 0.000280        &~~~~~ 0.000732            &~~~~~ 0.000368       &~~~~~-0.000088       &~~~~~ 0.000094 \\
\hline
~~~~ $A~(km)$      &~~~~~ 8.754660      &~~~~~ 4.339340      &~~~~~ 7.402050        &~~~~~-21.98830        &~~~~~ 24.58420 \\
\hline
~~~~ $B~(km)$      &~~~~~ 0.038589            &~~~~~ 0.040008           &~~~~~ 0.040052        &~~~~~ 0.037919        &~~~~~ 0.039431
  \\
\hline
\end{tabular}
\label{tab1}
\end{table}
\section{Physical Characteristics of the Compact Stars Models}
Here, we discuss the graphical behavior of the considered compact stars in the context of $f(R)$ gravity models. We studied the graphical illustration of energy density, pressure components, stability and equilibrium condition, energy conditions, mass function, adiabatic index, etc. For plotting the graphs, we consider the same values of constants mentioned in Tab. \ref{tab1}. Moreover, for Model-1, we fix the constant parameter $\alpha=0.1$ and the variation of $\beta,$ for different compact stars are given as: $\beta=1.5$ for $S_{1}$, $\beta=1$ for $S_{2}$, $\beta=1$ for $S_{3}$, $\beta=1.5$ for $S_{4}$ and $\beta=1$ for $S_{5}$. While for Model-2, we fix the constant parameters $\mu=5$ and $n=3.$
\subsection{Energy Density and Pressure Movement}
\begin{figure}[h!]
\begin{tabular}{cccc}
\epsfig{file=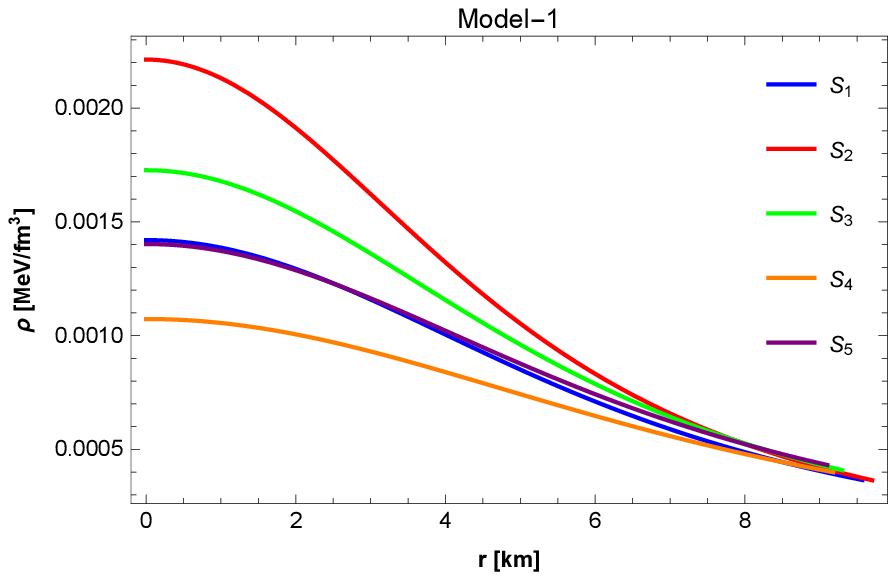,width=0.33\linewidth} &
\epsfig{file=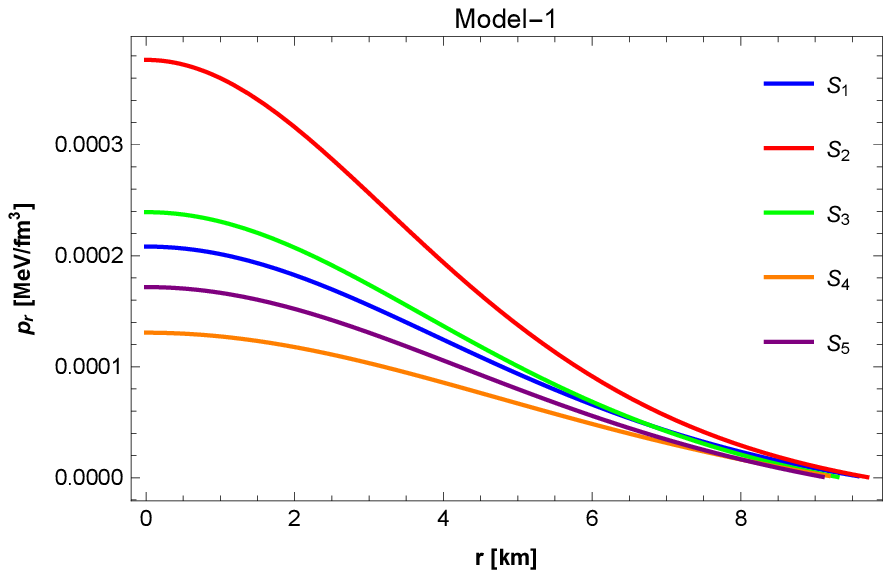,width=0.33\linewidth} &
\epsfig{file=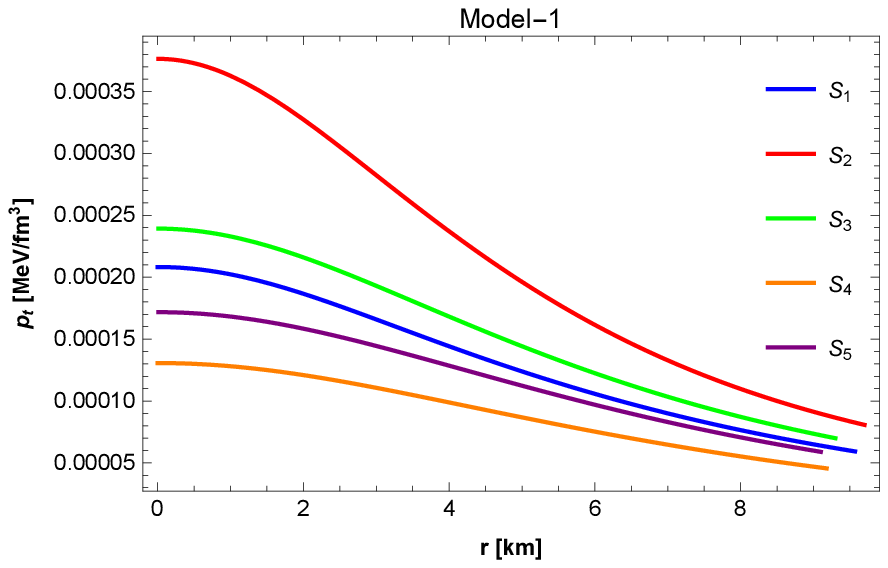,width=0.33\linewidth} &
\end{tabular}
\caption{{Behavior of $\rho$, $p_{r}$ and $p_{t}$ for Model-1}.}
\label{Fig:2}
\end{figure}

\begin{figure}[h!]
\begin{tabular}{cccc}
\epsfig{file=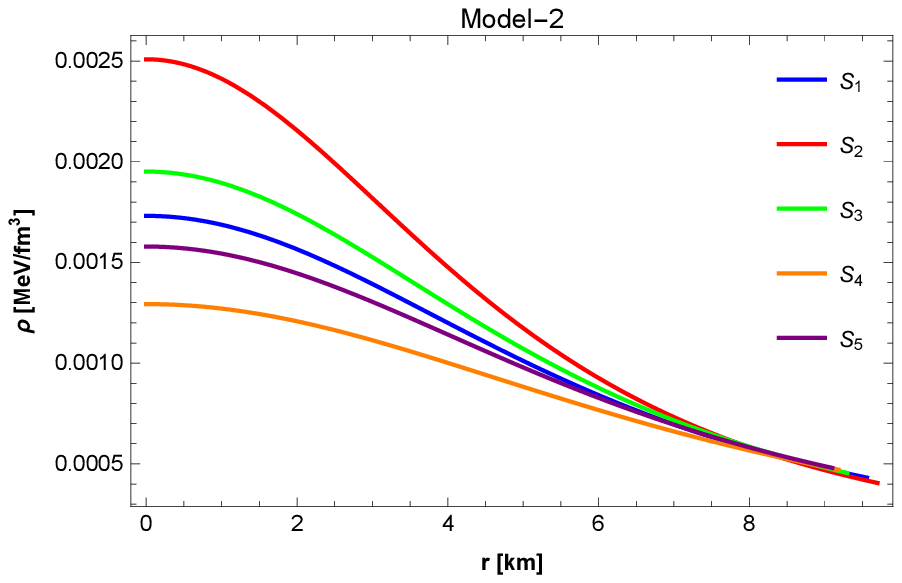,width=0.33\linewidth} &
\epsfig{file=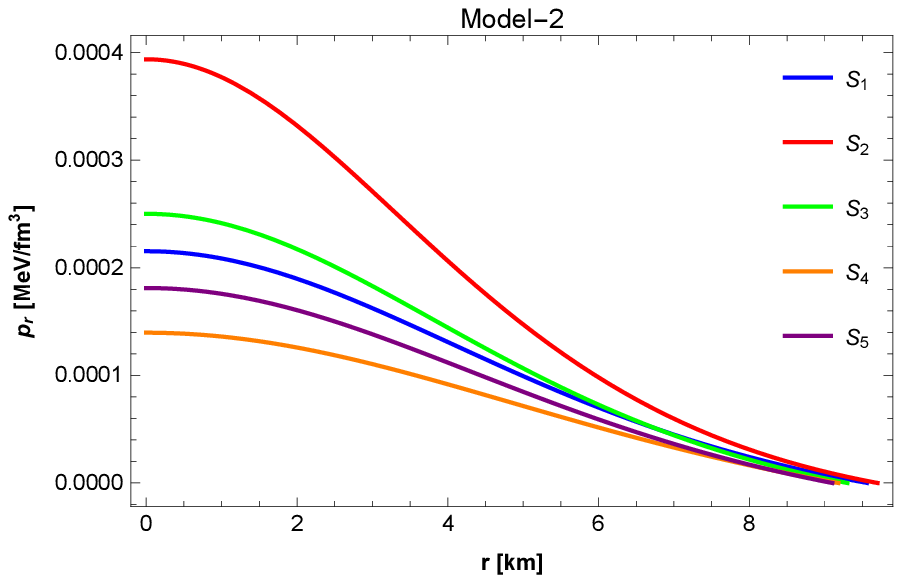,width=0.33\linewidth} &
\epsfig{file=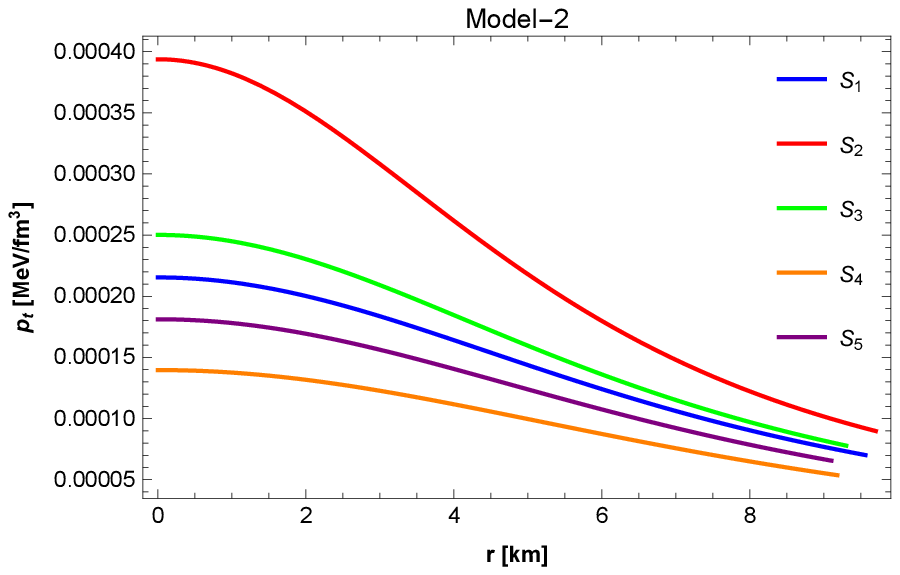,width=0.33\linewidth} &
\end{tabular}
\caption{{Behavior of $\rho$, $p_{r}$ and $p_{t}$ for Model-2}.}
\label{Fig:3}
\end{figure}

\begin{figure}[h!]
\begin{tabular}{cccc}
\epsfig{file=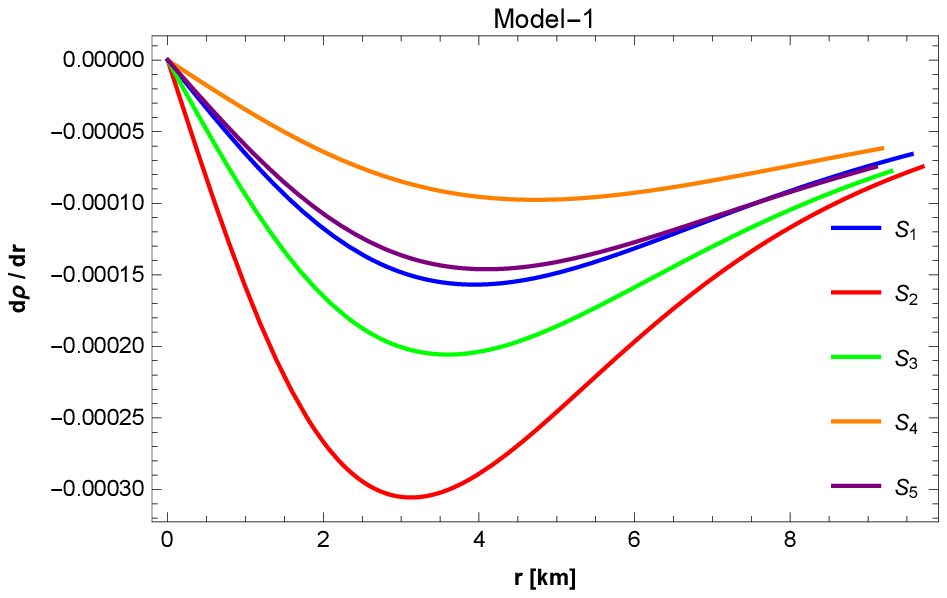,width=0.33\linewidth} &
\epsfig{file=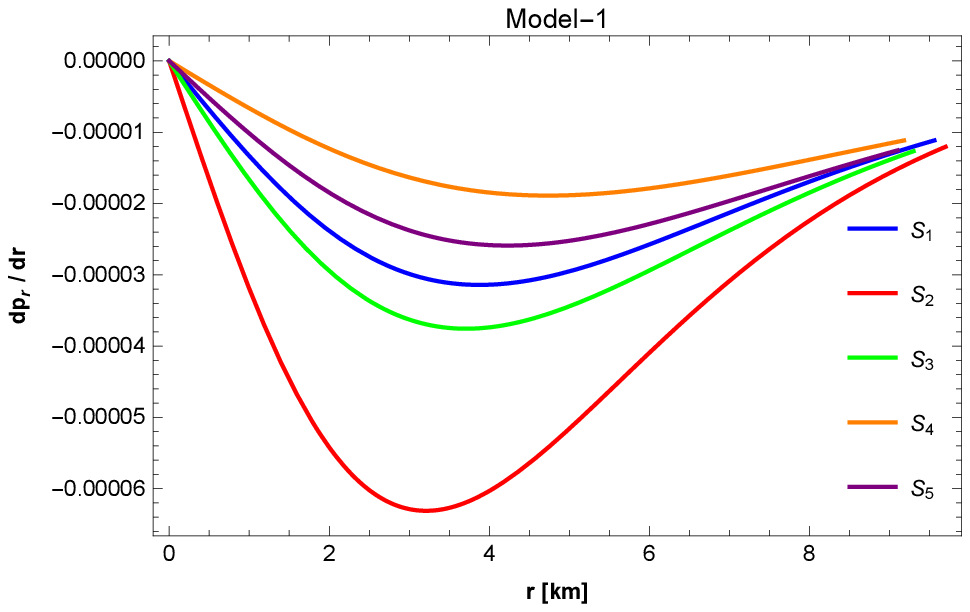,width=0.33\linewidth} &
\epsfig{file=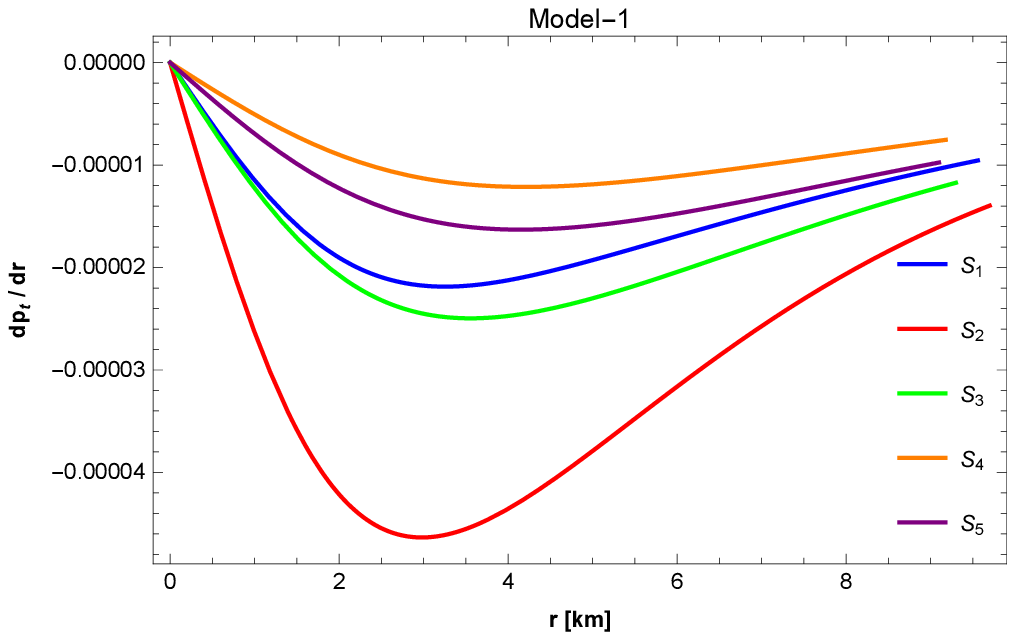,width=0.33\linewidth} &
\end{tabular}
\caption{{Variation of $\frac{d\rho}{dr},$ $\frac{dp_{r}}{dr}$ and $\frac{dp_{t}}{dr}$ for Model-1}.}
\label{Fig:4}
\end{figure}

\begin{figure}[h!]
\begin{tabular}{cccc}
\epsfig{file=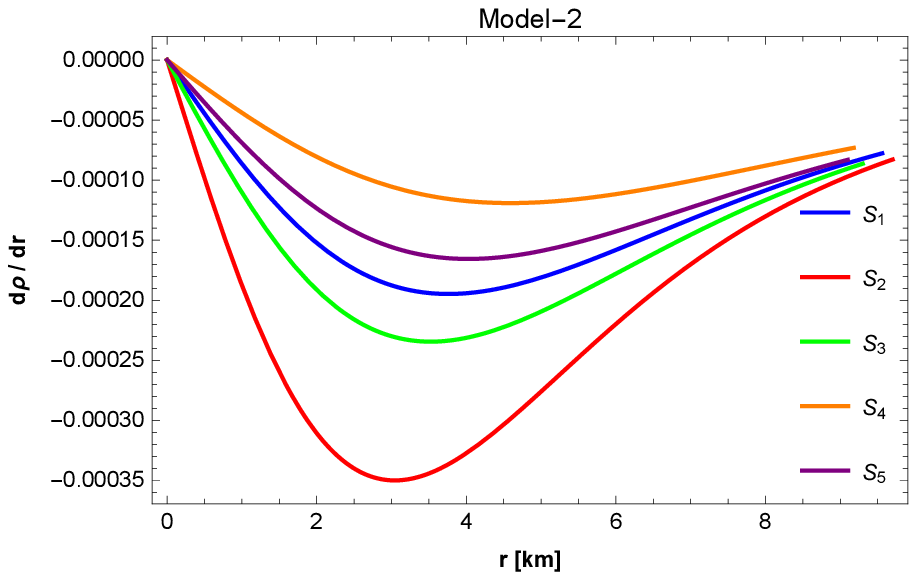,width=0.33\linewidth} &
\epsfig{file=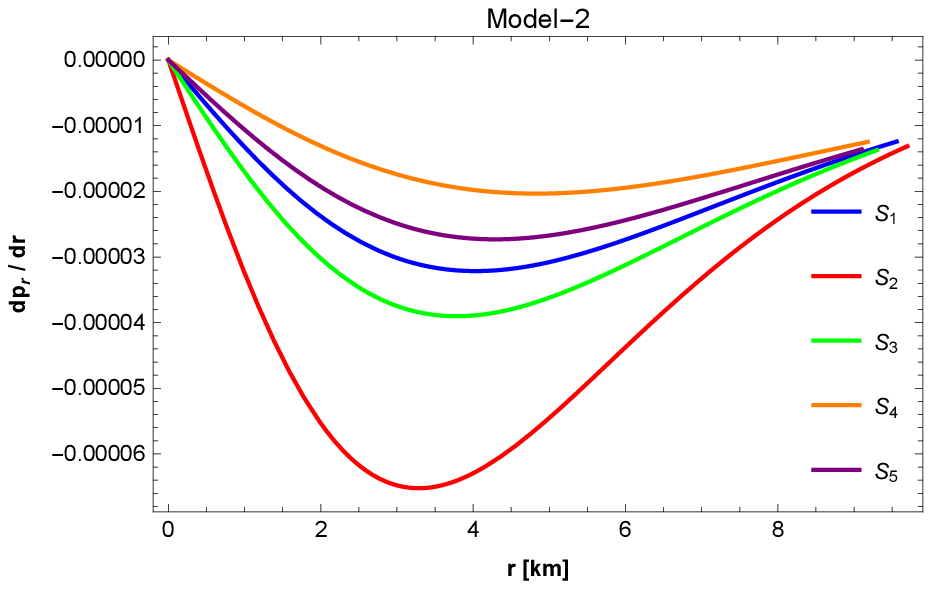,width=0.33\linewidth} &
\epsfig{file=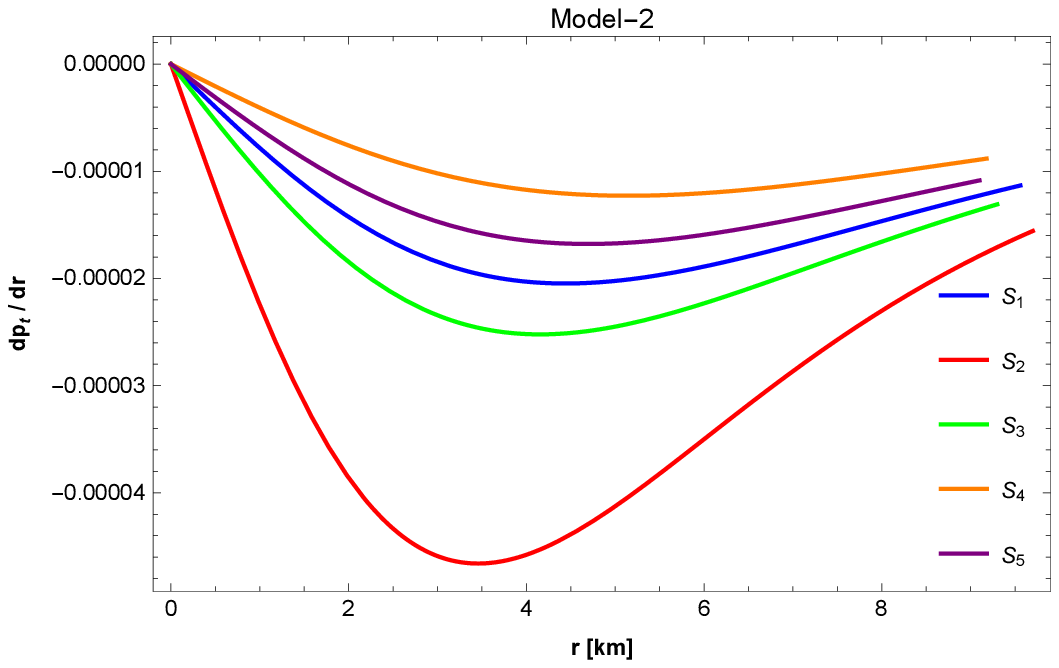,width=0.33\linewidth} &
\end{tabular}
\caption{{Variation of $\frac{d\rho}{dr},$ $\frac{dp_{r}}{dr}$ and $\frac{dp_{t}}{dr}$ for Model-2}.}
\label{Fig:5}
\end{figure}

In this portion, we represent the graphical response of $\rho,$ $p_{r}$ and $p_{t}$ for considered $f(R)$ theory of gravity models. The graphical behavior of the preceding aspects are represented in Fig. $\ref{Fig:2}$ and Fig. $\ref{Fig:3}.$ Furthermore, we also present the graphical behavior of gradient of $\rho,$ $p_{r}$ and $p_{t}$. There graphs have been shown in Fig. $\ref{Fig:4}$ and Fig. $\ref{Fig:5}.$\\
It is clearly viable from Fig. $\ref{Fig:2}$ and Fig. $\ref{Fig:3}$ that $\rho,$ $p_{r}$ and $p_{t}$ plots are positive and finite. Further, one can easily observed that the illustration of an energy density and pressure components attain the highest values at compact star center and decreases toward surface boundary.\\

The gradient of $\rho,$ $p_{r}$ and $p_{t}$ presented negative, which can be seen from Fig. $\ref{Fig:4}$ and Fig. $\ref{Fig:5}.$ These facts show the high compactness nature of the compact stars.
\subsection{Anisotropy}
\begin{figure}[h!]
\begin{tabular}{cccc}
\epsfig{file=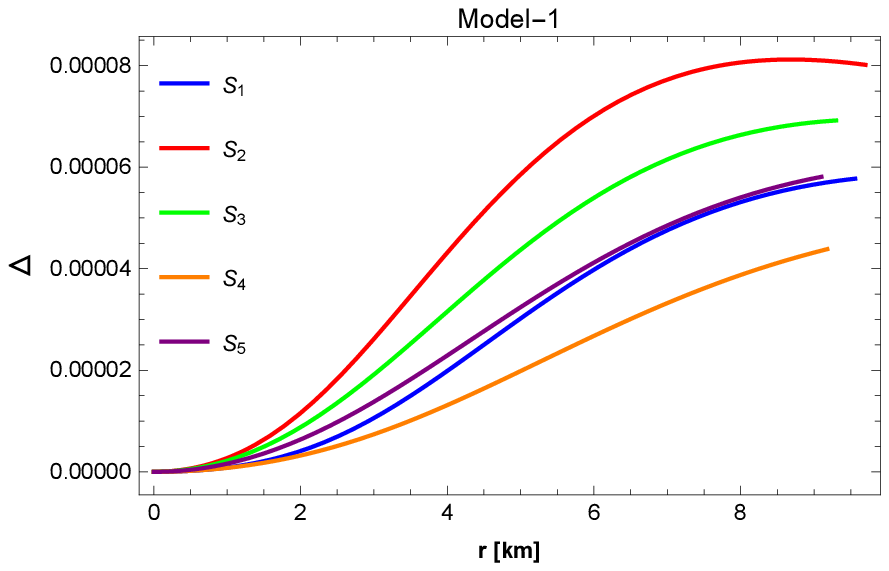,width=0.38\linewidth} &
\epsfig{file=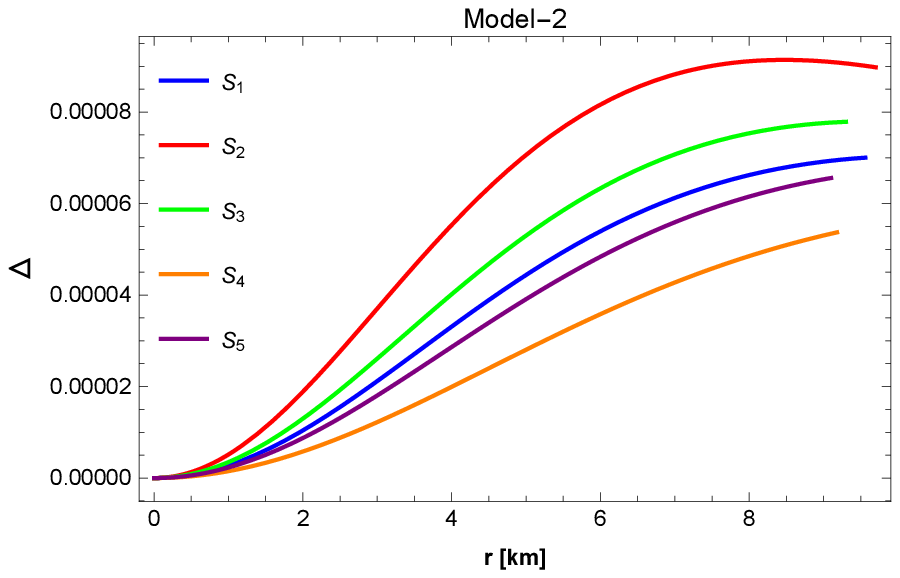,width=0.38\linewidth} &
\end{tabular}
\caption{{Evolution of anisotropy}.}
\label{Fig:6}
\end{figure}

Next, we discuss the graphical behavior of anisotropy parameter \cite{Harko}, symbolized by $\Delta$ and shown in Eq. (\ref{6}). For the existence of the compact stars the anisotropic force must be repulsive \cite{A2}. From Fig. $\ref{Fig:6}$, it is worthwhile to mention here that in our study $\Delta>0$, which means that our models exhibit repulsive nature.

\subsection{Energy Conditions}
\begin{figure}[h!]
\begin{tabular}{cccc}
\epsfig{file=D1.eps,width=0.33\linewidth} &
\epsfig{file=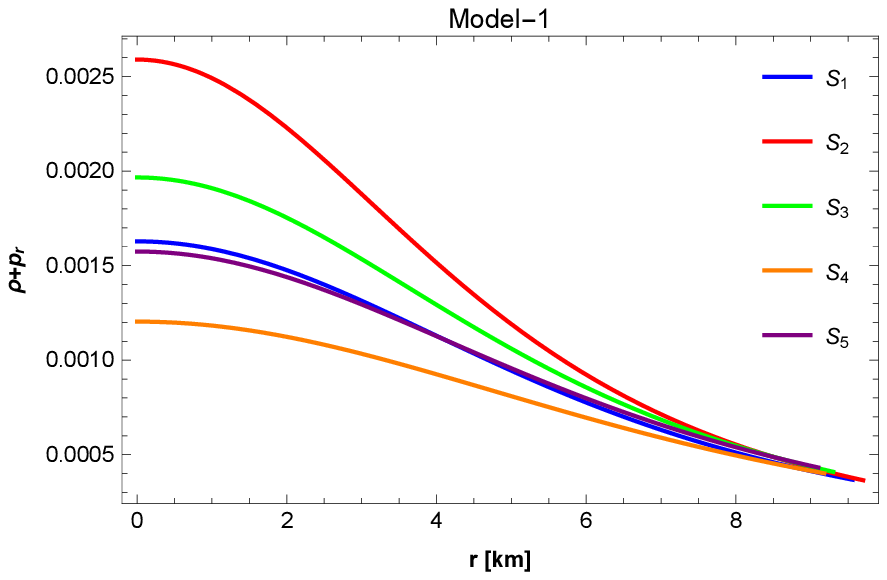,width=0.33\linewidth} &
\epsfig{file=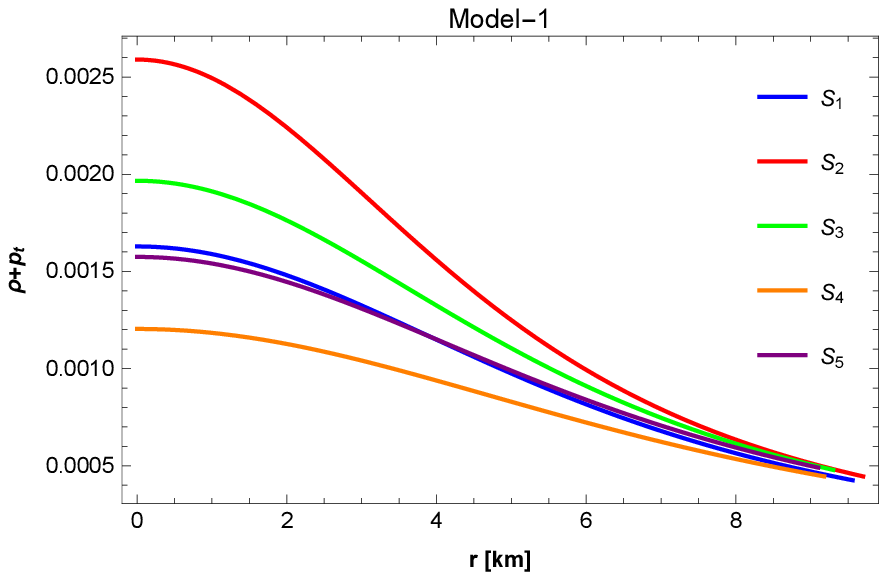,width=0.33\linewidth} &\\
\epsfig{file=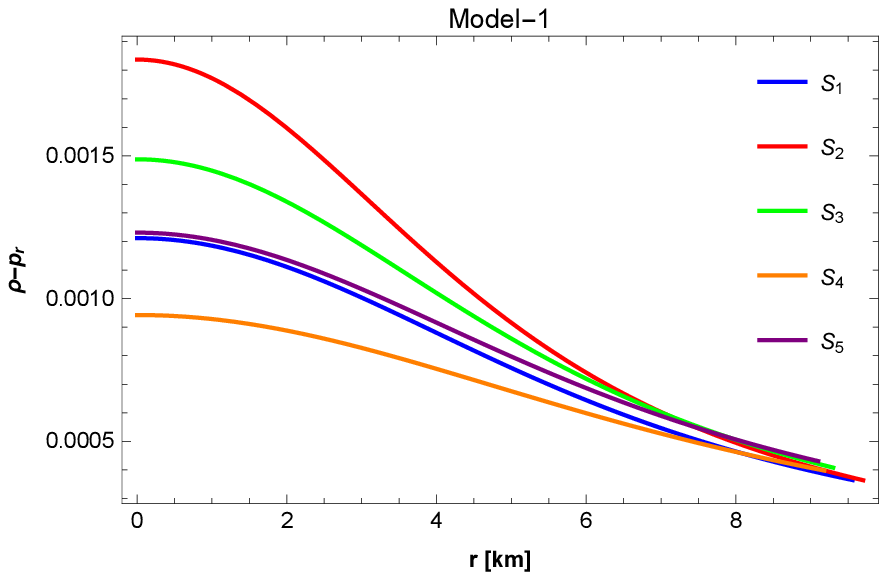,width=0.33\linewidth} &
\epsfig{file=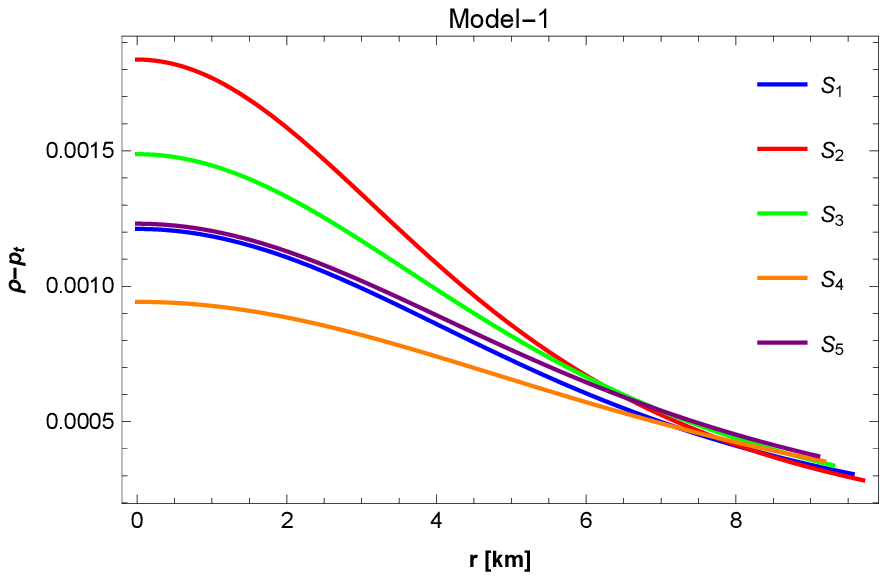,width=0.33\linewidth} &
\epsfig{file=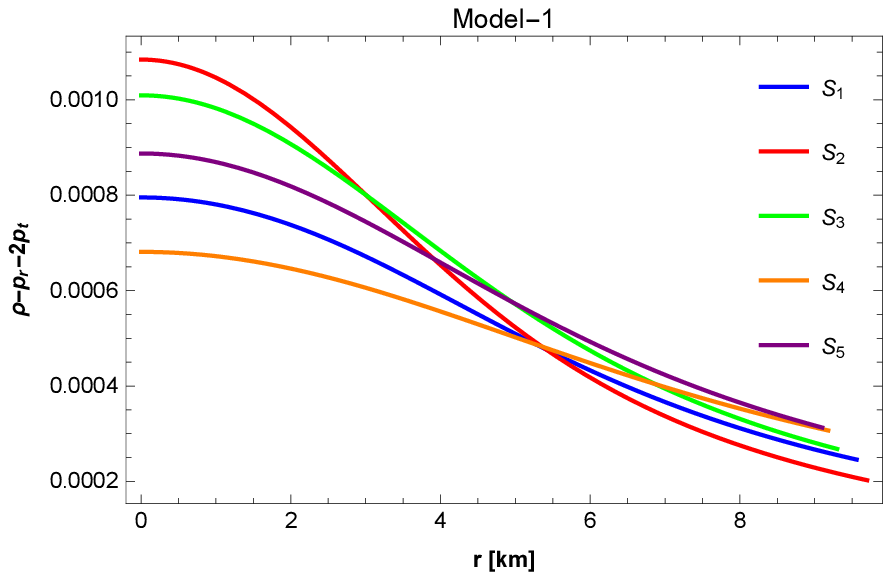,width=0.33\linewidth} &
\end{tabular}
\caption{{Graphs of energy conditions for Model-1}.}
\label{Fig:7}
\end{figure}

\begin{figure}[h!]
\begin{tabular}{cccc}
\epsfig{file=D2.eps,width=0.33\linewidth} &
\epsfig{file=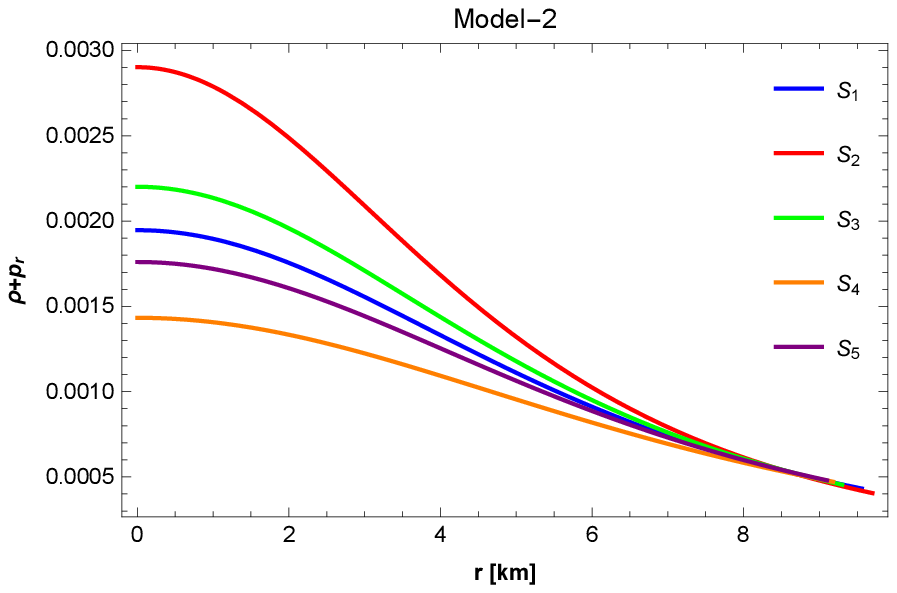,width=0.33\linewidth} &
\epsfig{file=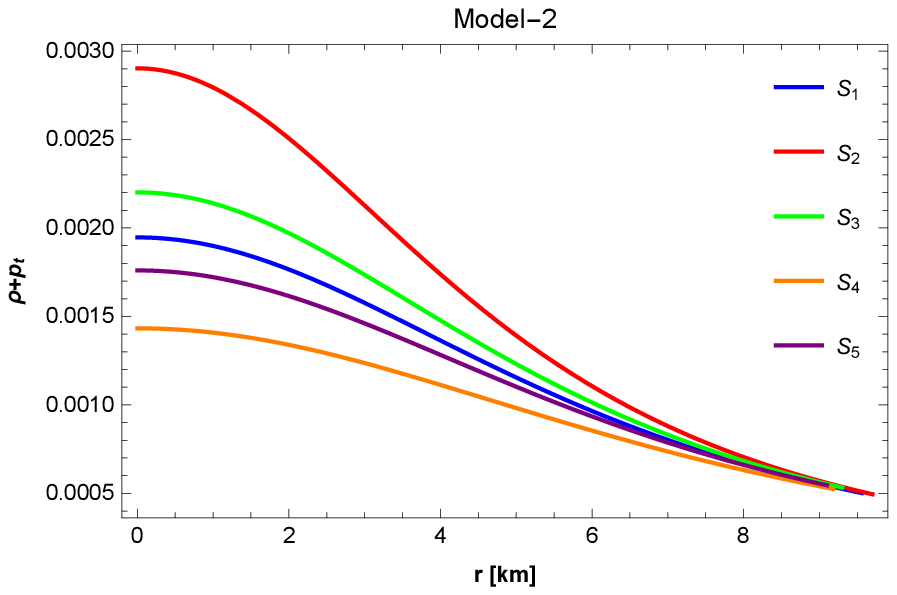,width=0.33\linewidth} &\\
\epsfig{file=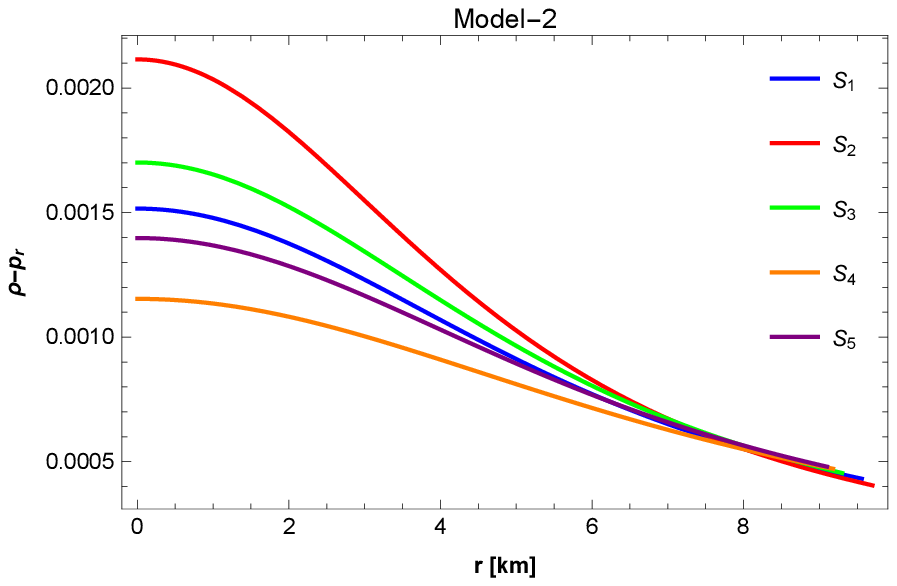,width=0.33\linewidth} &
\epsfig{file=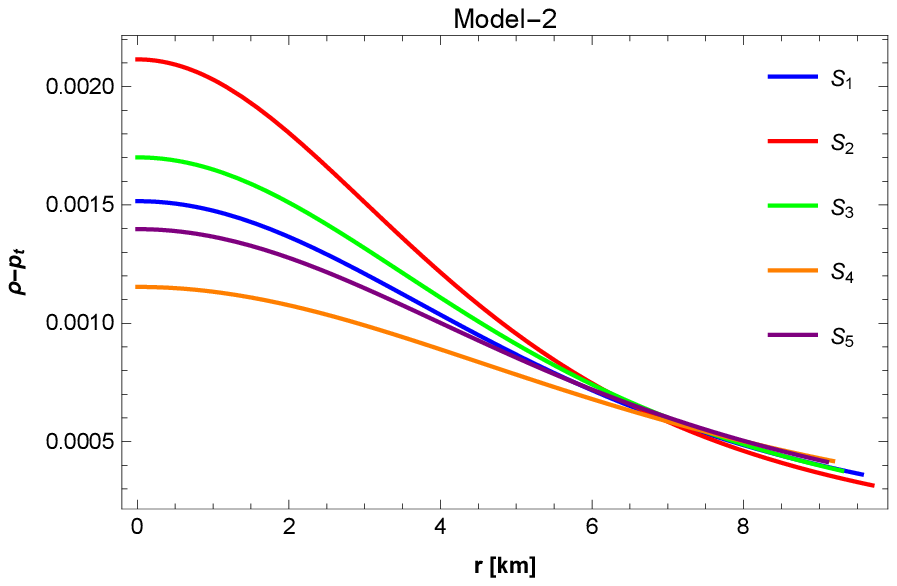,width=0.33\linewidth} &
\epsfig{file=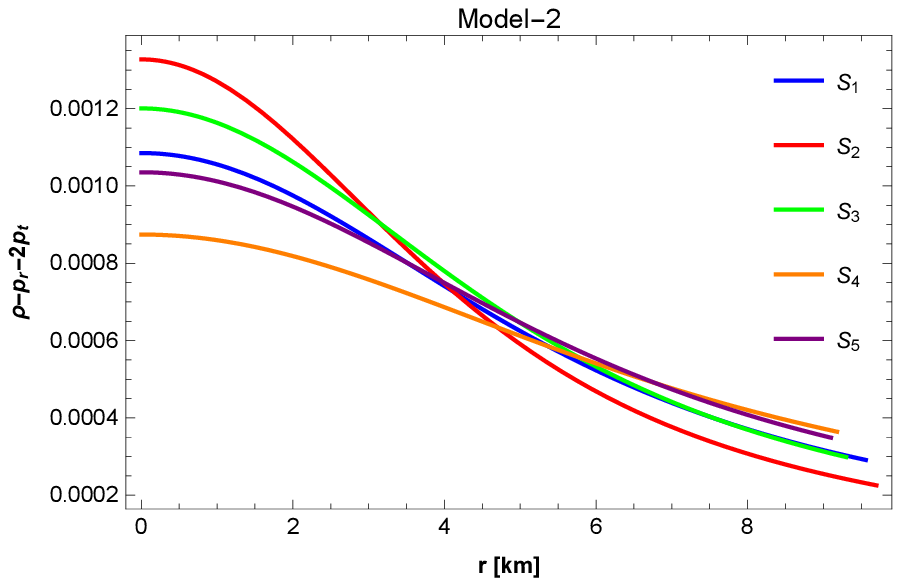,width=0.33\linewidth} &
\end{tabular}
\caption{{Graphs of energy conditions for Model-2}.}
\label{Fig:8}
\end{figure}

In the existence of the compact stars, energy conditions plays a significant role. These energy bonds \cite{You} are categories as null energy, weak energy, strong energy and dominant energy conditions symbolized by NEC, WEC, SEC and DEC, respectively and defined as
\begin{equation*}
  NEC:  \rho+p_{r} \geq 0,\rho+p_{t},~~~~~~WEC:\rho \geq 0,\rho+p_{r} \geq 0,\rho+p_{t} \geq 0,~~~~~~SEC:\rho-p_{r} -2p_{t}\geq 0,~~~~~~DEC:\rho \geq 0,\rho\pm p_{r} \geq 0,\rho\pm p_{t} \geq 0. \\
  \end{equation*}
It has been observed from Fig. $\ref{Fig:7}$ and Fig. $\ref{Fig:8}$ that all energy bonds have been well satisfied for our proposed models. 

\subsection{Equilibrium Condition}
\begin{figure}[h!]
\begin{tabular}{cccc}
\epsfig{file=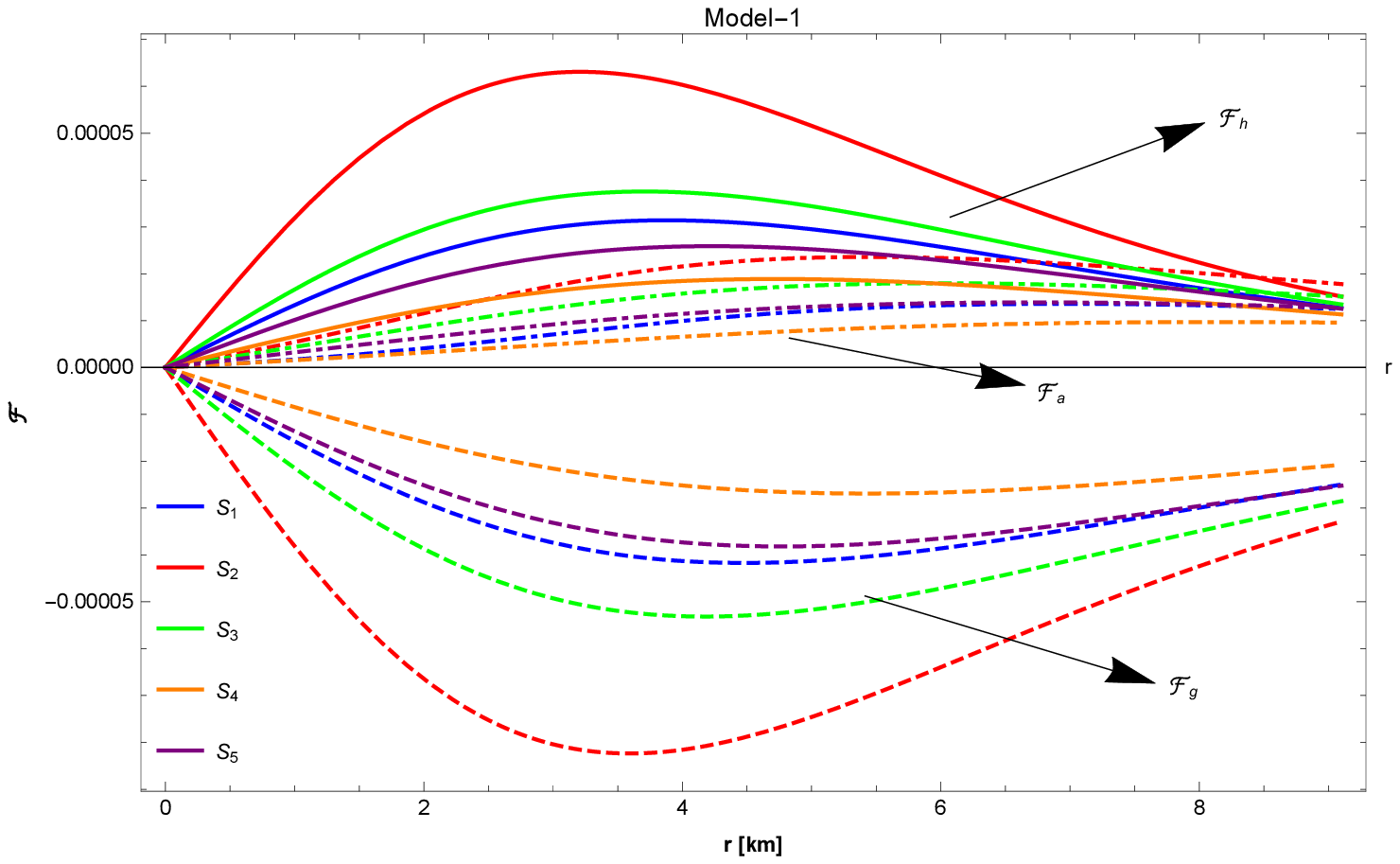,width=0.38\linewidth} &
\epsfig{file=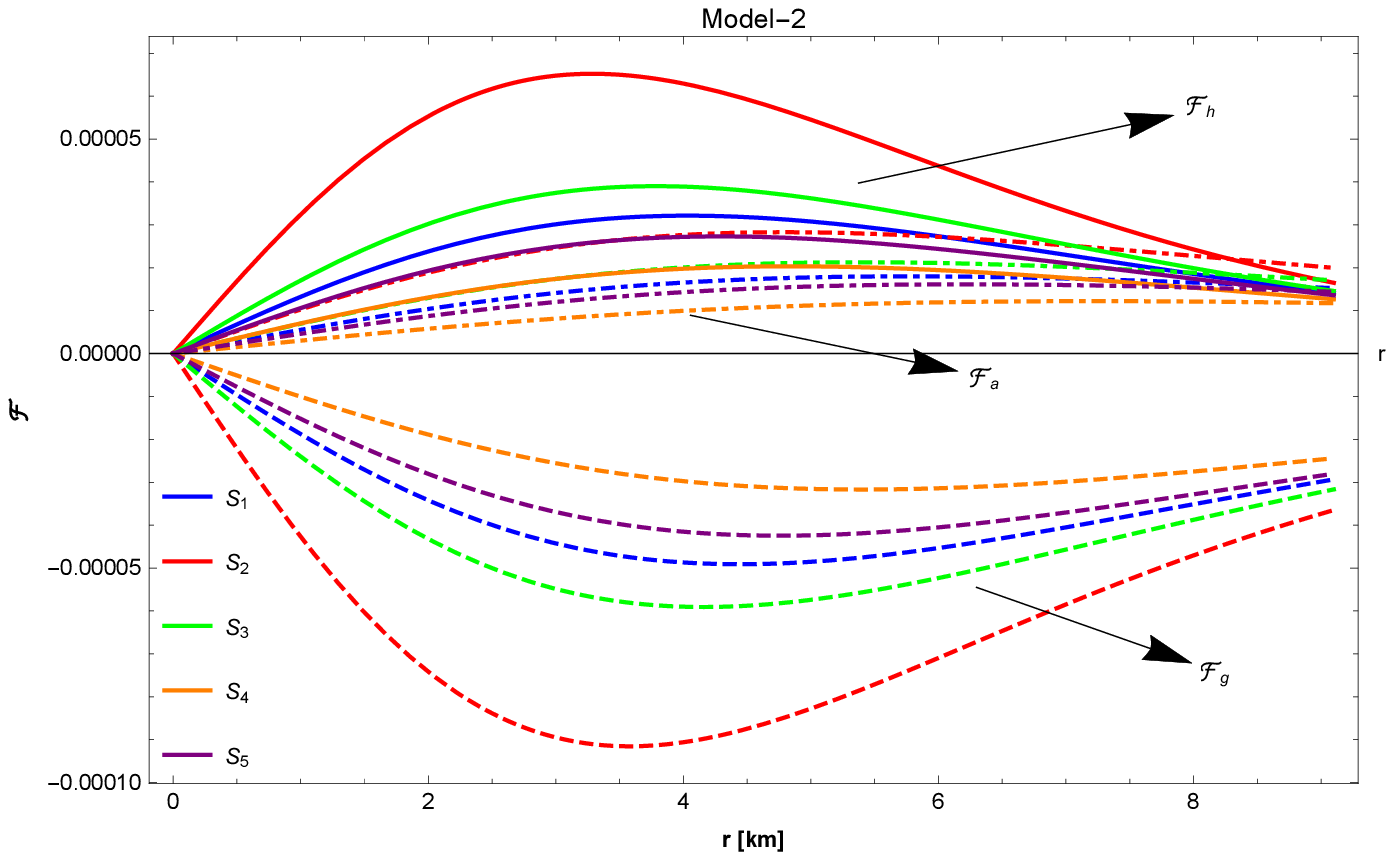,width=0.38\linewidth} &
\end{tabular}
\caption{{Behavior of $\mathcal{F}_{h}$, $\mathcal{F}_{g}$ and $\mathcal{F}_{a}$ for our proposed models}.}
\label{Fig:9}
\end{figure}

In this subsection, we explore the equilibrium stability of the stellar model under the presence of gravitational force, hydrostatics force and anisotropic force. According to the TOV equation \cite{Tolman,Oppen}, the equilibrium condition between all these forces can be described as
\begin{equation}\label{19}
 \frac{ M_{G}(r)(\rho+p_{r})}{r}e^{\frac{\lambda-\nu}{2}}+\frac{dp_{r}}{dr}-\frac{2}{r}(p_{t}-p_{r})=0.
\end{equation}
 The effective gravitational mass $ M_{G}(r)$ is defined as
  \begin{equation}\label{20}
    M_{G}(r)=\frac{1}{2}\nu^{'}e^{\frac{\nu-\lambda}{2}}.
  \end{equation}
Putting value of $ M_{G}(r)$ in Eq. (\ref{19}), we get
\begin{equation}\label{21}
 \frac{\nu^{'}}{r}(\rho+p_{r})+\frac{dp_{r}}{dr}-\frac{2}{r}(p_{t}-p_{r})=0,
\end{equation}
where, $ \mathcal{F}_{g} = -\frac{\nu^{'}}{r}(\rho+p_{r}),~\mathcal{F}_{h} = -\frac{dp_{r}}{dr},~ \mathcal{F}_{a} = \frac{2}{r}\Delta.$
Here, $\mathcal{F}_{g}$ designates gravitational force, $\mathcal{F}_{h}$ designates hydrostatic force and $\mathcal{F}_{a}$ designates anisotropic force. For viable $f(R)$ gravity models, all three forces necessarily express the balancing trend, i.e. for a system to be in equilibrium the joint sum of three forces should be exactly equal to zero. Consequently, the Eq. (\ref{21}) yield
$$\mathcal{F}_{g}+\mathcal{F}_{h}+\mathcal{F}_{a}=0.$$
From Fig. $\ref{Fig:9}$, it can be easily seen that all forces are satisfied the necessary condition of equilibrium.

\subsection{Equation of State}
In this portion, we compute the equations of state (EoS) for both radial and transverse pressures. The EoS are presented by two ratios, which are mentioned as
\begin{equation}\label{22}
   w_{r} = \frac{p_{r}}{\rho},~~~~~~~~w_{t} =  \frac{p_{t}}{\rho}.
\end{equation}
\begin{figure}[h!]
\begin{tabular}{cccc}
\epsfig{file=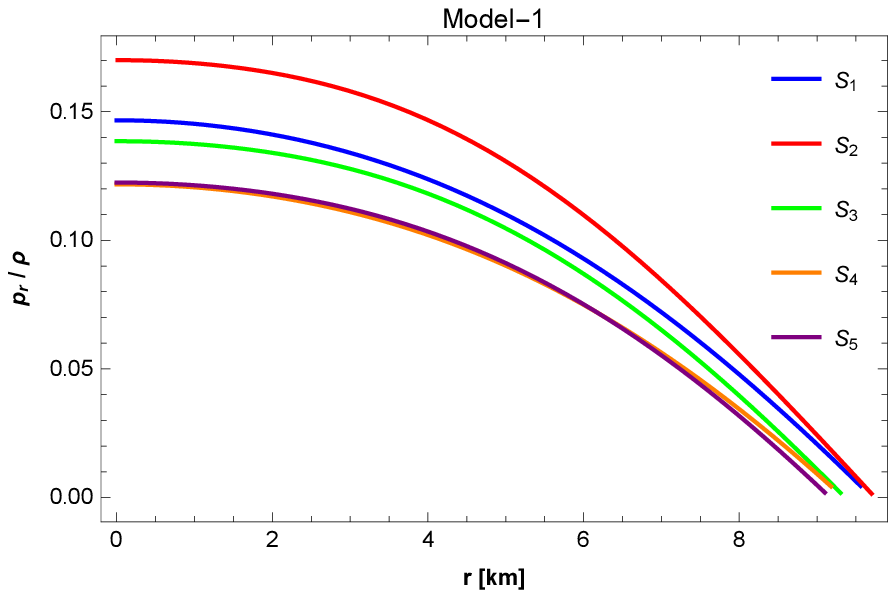,width=0.38\linewidth} &
\epsfig{file=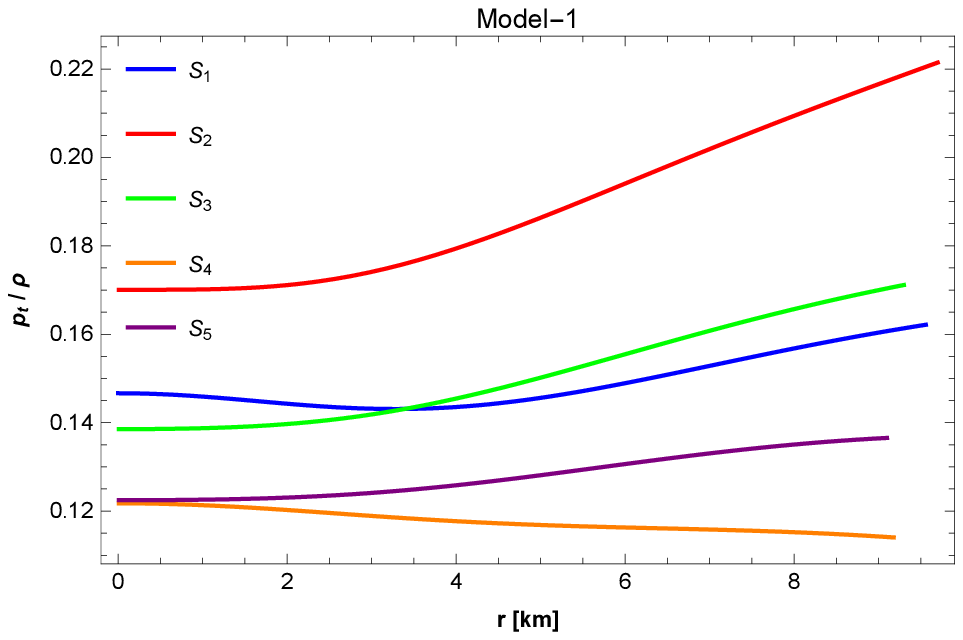,width=0.38\linewidth} &
\end{tabular}
\caption{{Evolution of $w_{r}$ and $w_{t}$ for Model-1}.}
\label{Fig:10}
\end{figure}

\begin{figure}[h!]
\begin{tabular}{cccc}
\epsfig{file=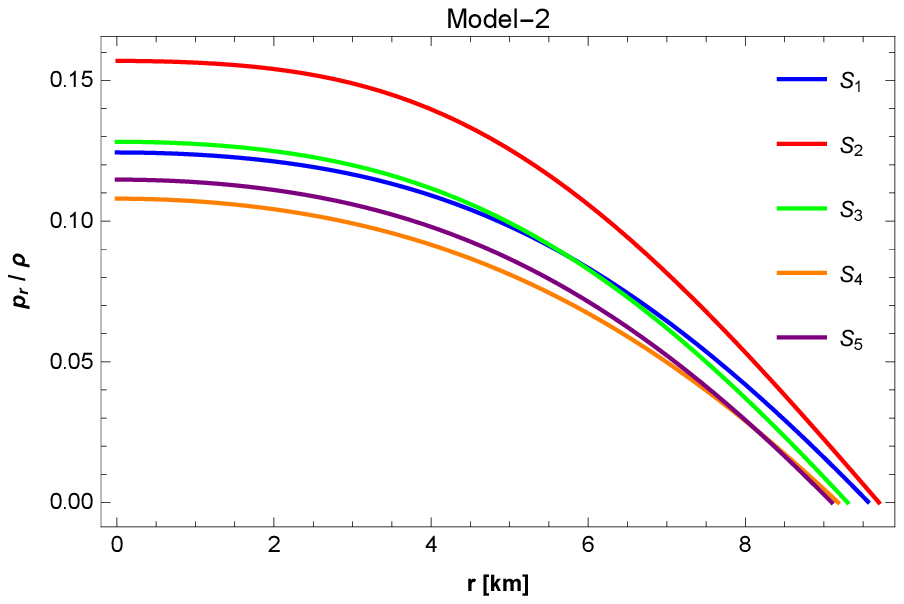,width=0.38\linewidth} &
\epsfig{file=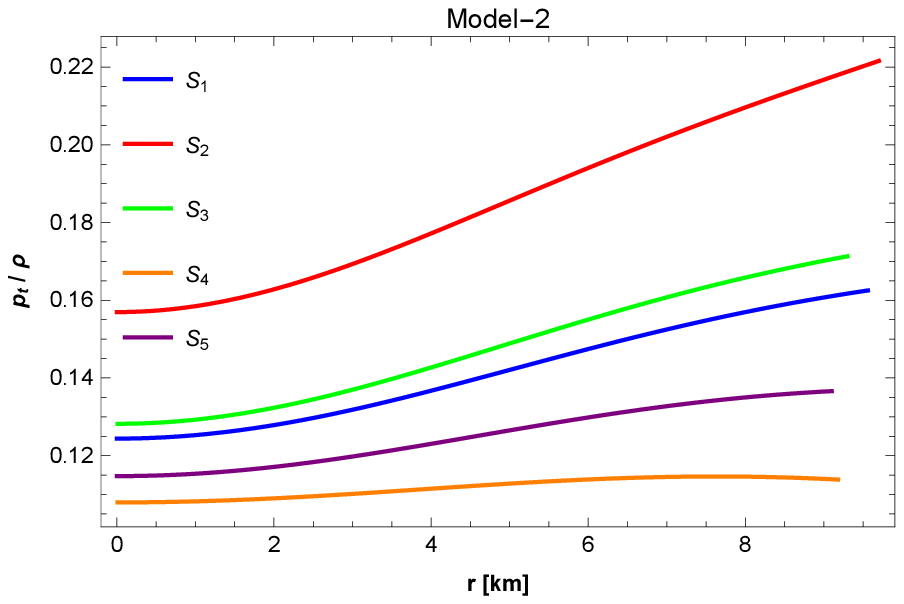,width=0.38\linewidth} &
\end{tabular}
\caption{{Evolution of $w_{r}$ and $w_{t}$ for Model-2}.}
\label{Fig:11}
\end{figure}

From the graphical illustration shown in Fig. $\ref{Fig:10}$ and Fig. $\ref{Fig:11}$, it can be noticed that $0<w_{r},~w_{t}<1.$ Hence, we can conclude that the variational responses of EoS are sufficient for a relativistic $f(R)$ gravity model.
\subsection{Causality Condition}
\begin{figure}[h!]
\begin{tabular}{cccc}
\epsfig{file=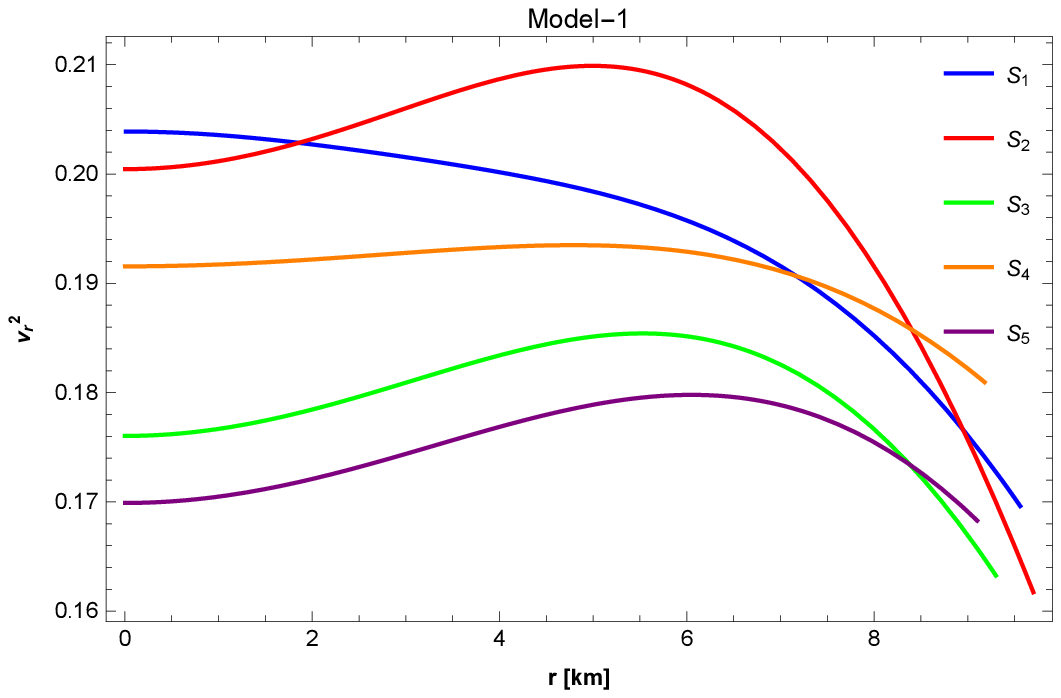,width=0.33\linewidth} &
\epsfig{file=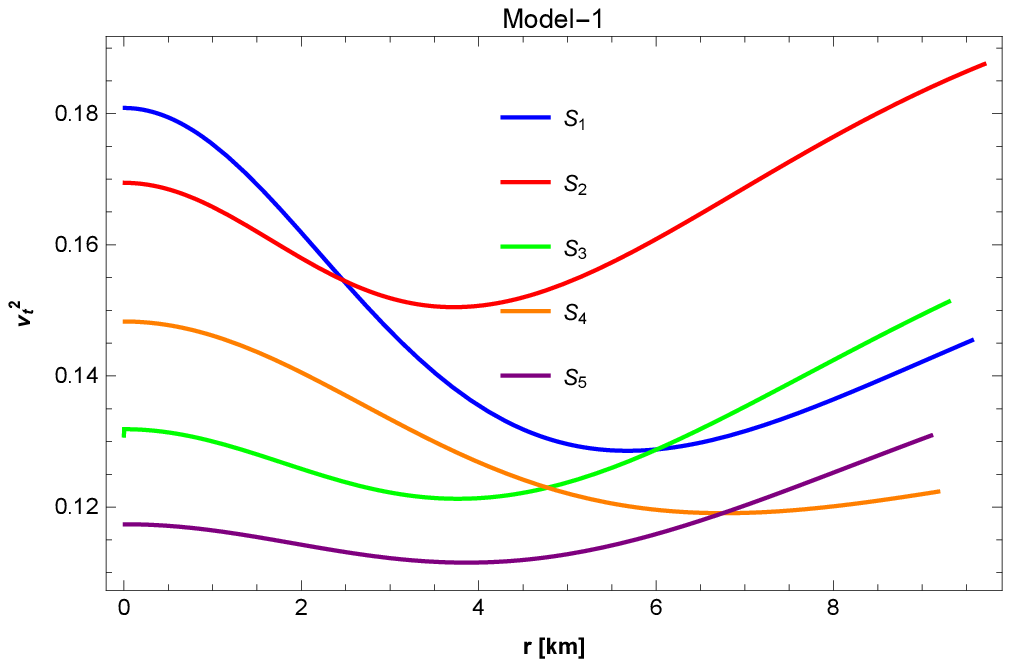,width=0.33\linewidth} &
\epsfig{file=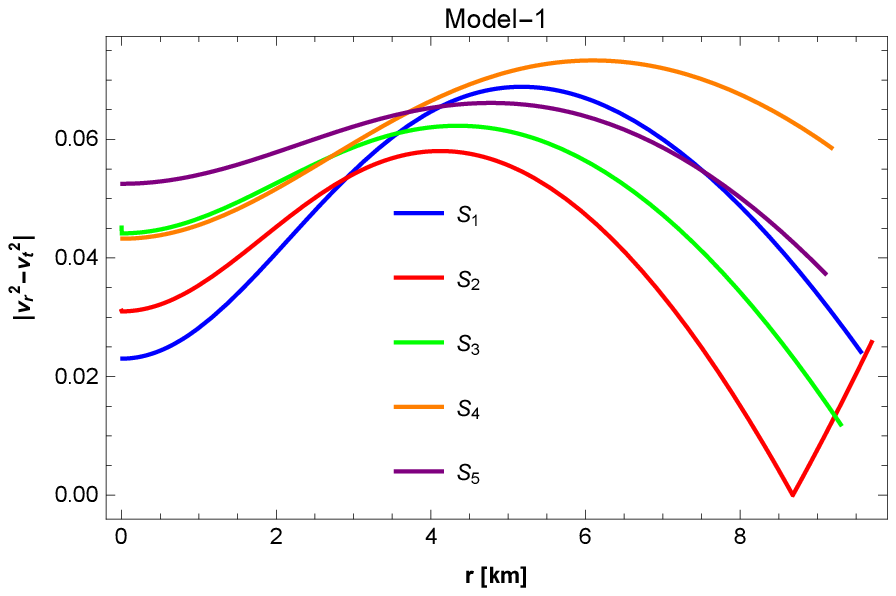,width=0.33\linewidth} &
\end{tabular}
\caption{{Behavior of ${v_{r}}^{2},$ ${v_{t}}^{2}$ and $|{v_{r}}^{2}-{v_{t}}^{2}|$ for $f(R)$ gravity Model-1}.}
\label{Fig:12}
\end{figure}

\begin{figure}[h!]
\begin{tabular}{cccc}
\epsfig{file=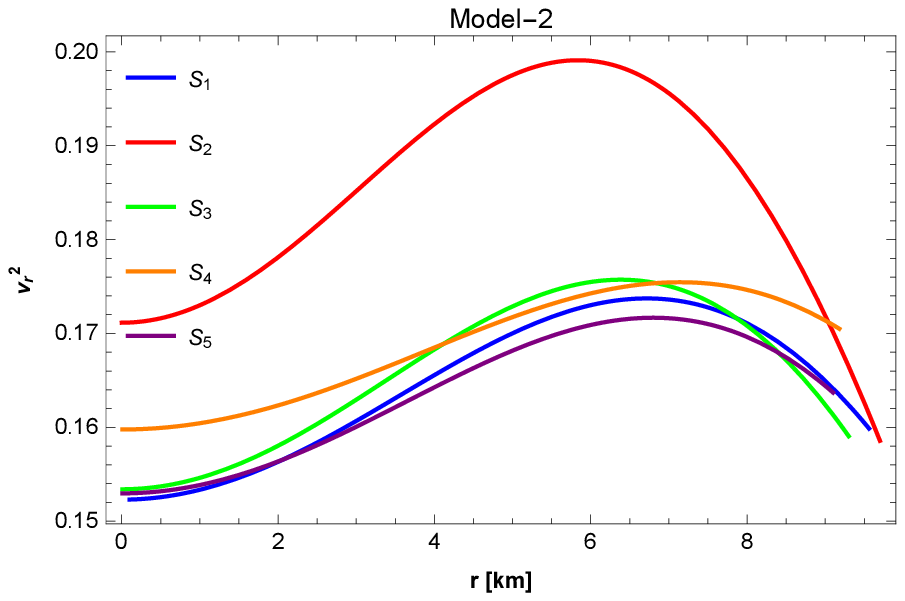,width=0.33\linewidth} &
\epsfig{file=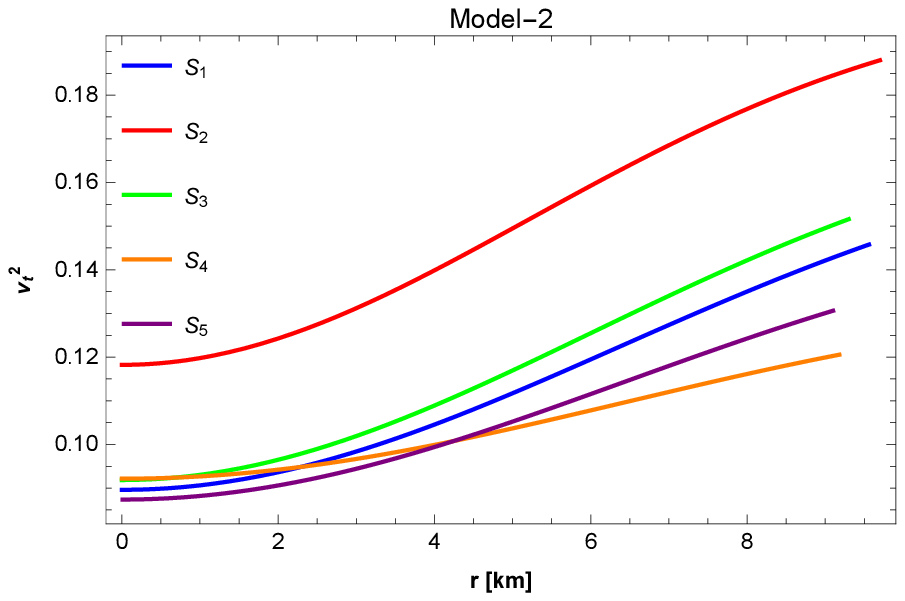,width=0.33\linewidth} &
\epsfig{file=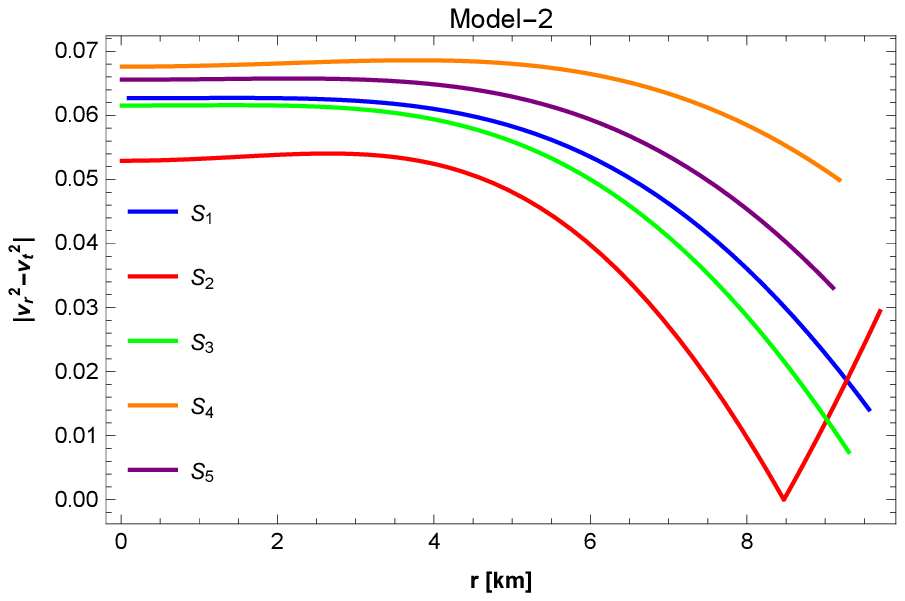,width=0.33\linewidth} &
\end{tabular}
\caption{{Behavior of ${v_{r}}^{2},$ ${v_{t}}^{2}$ and $|{v_{r}}^{2}-{v_{t}}^{2}|$ for $f(R)$ gravity Model-2}.}
\label{Fig:13}
\end{figure}

The velocity sound for radial pressure component ${v_{r}}^{2}$ and tangential pressure component ${v_{t}}^{2}$ are defined as
\begin{equation}\label{23}
  {v_{r}}^{2} = \frac{dp_{r}}{d\rho},~~~~~~~~{v_{t}}^{2} = \frac{dp_{t}}{d\rho}.
\end{equation}
To verify the stability of our chosen system, we use the Herrera concept \cite{Her} according to that ${v_{r}}^{2}$ and ${v_{t}}^{2}$ must lies within the range of $0$ and $1.$ Moreover, we also studied the graphical behavior of our chosen compact stars for the Andreasson condition \cite{And}, that is, $|{v_{r}}^{2}-{v_{t}}^{2}|\leq1.$ The graphical illustration presented in Fig. $\ref{Fig:12}$ and Fig. $\ref{Fig:13}$ confirm that the considered $f(R)$ gravity models are potentially stable and satisfied both causality condition and Herrera cracking concept.

\subsection{Mass Function, Compactness Factor and Redshift Analysis}
\begin{figure}[h!]
\begin{tabular}{cccc}
\epsfig{file=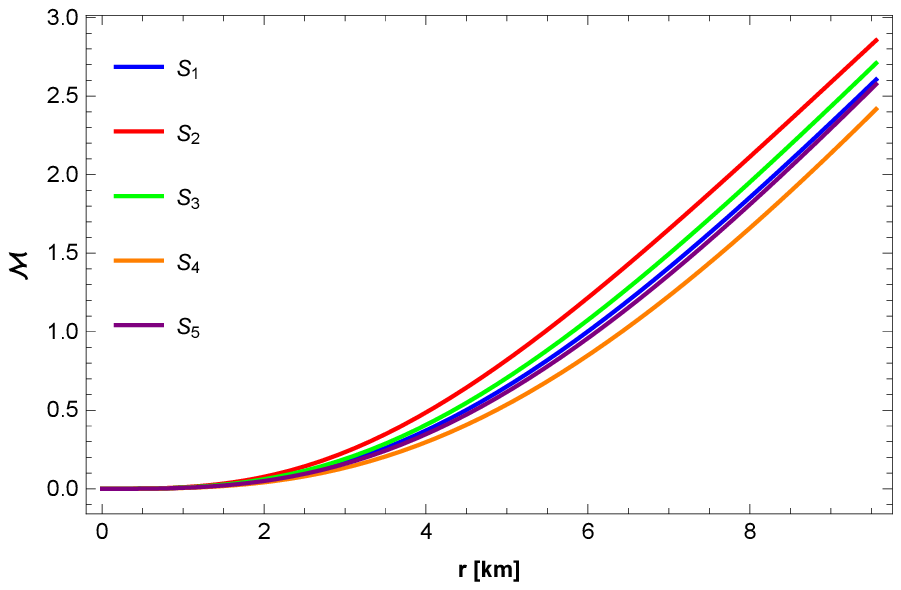,width=0.33\linewidth} &
\epsfig{file=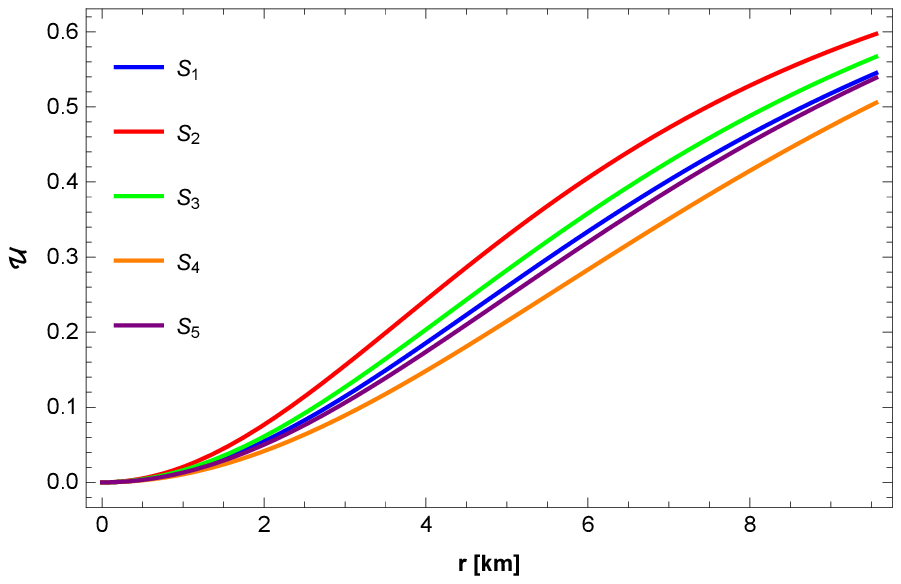,width=0.33\linewidth} &
\epsfig{file=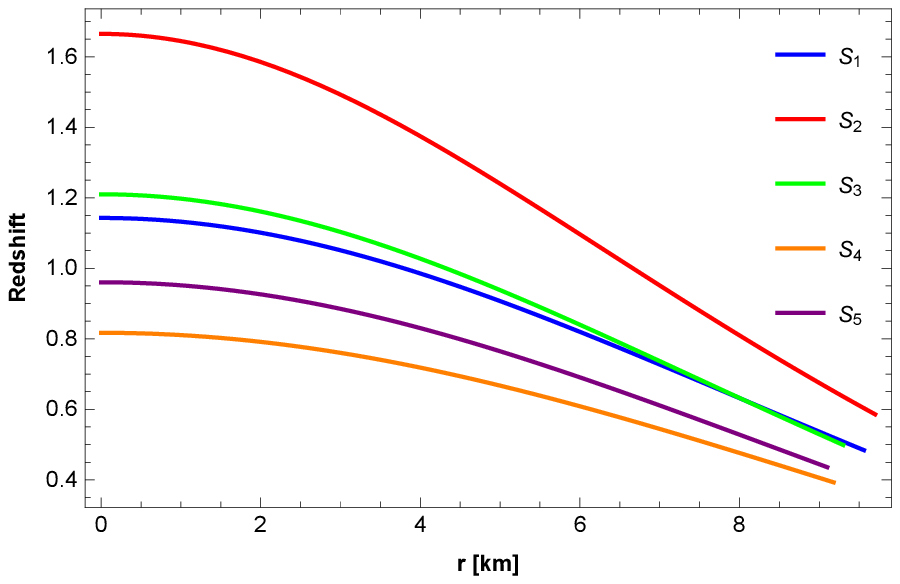,width=0.33\linewidth} &
\end{tabular}
\caption{{Behavior of Mass function (left panel), Compactness Factor (middle panel) and Redshift (right panel)}.}
\label{Fig:14}
\end{figure}

Using the metric potential $g_{rr}^{-}=g_{rr}^{+}$ the mass function \cite{Bhar4} is obtained as
\begin{equation}\label{24}
  \mathcal{M}(r)= \frac{a^{2}r^{3}}{2[(1+br^{2})^{2}+a^{2}r^{2}]}.
\end{equation}\label{25}
Next, we calculate the compactness factor \cite{Maha} and redshift \cite{Harko} which are given by the following mathematical relation
\begin{equation}
\mathcal{U}(r)=\frac{2\mathcal{M}(r)}{r}=\frac{a^{2}r^{2}}{[(1+br^{2})^{2}+a^{2}r^{2}]},
\end{equation}
\begin{equation}
  \mathcal{Z} = e^{-\frac{\nu}{2}}-1.
\end{equation}
The graphical response of mass function, compactness parameter and redshift are given in Fig. $\ref{Fig:14}.$ The behavior of mass function satisfies the Buchdahl \cite{Buchdahl} and Bondi \cite{Bon} limit, that is, $\frac{2M}{R}<\frac{8}{9}.$ Furthermore, the mass function and compactness parameter shown monotonically increasing plots whereas the graphical behavior of redshift is monotonically decreasing. Thus, we conclude that all the three graphs are stable for our $f(R)$ gravity models.

\subsection{Adiabatic Index}
\begin{figure}[h!]
\begin{tabular}{cccc}
\epsfig{file=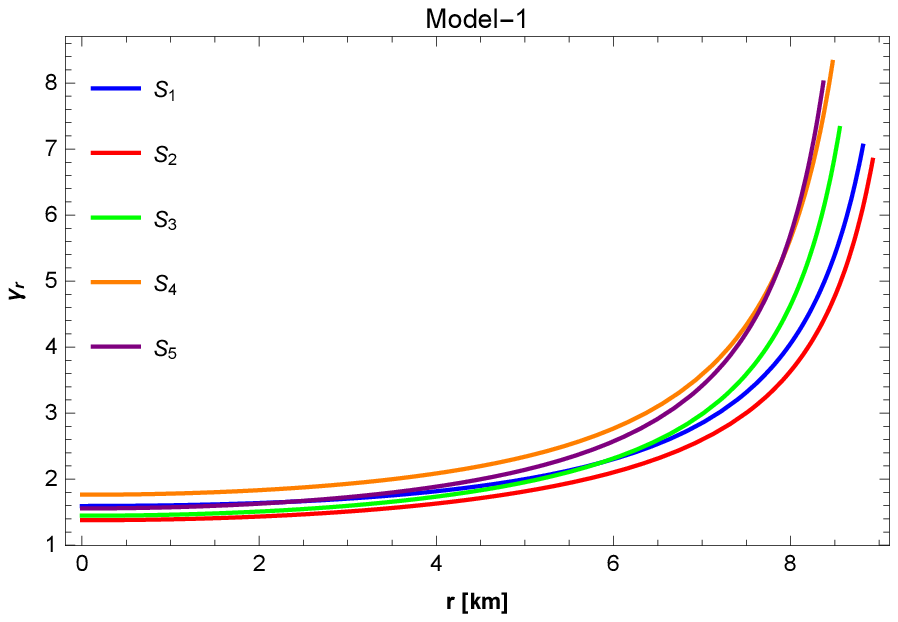,width=0.38\linewidth} &
\epsfig{file=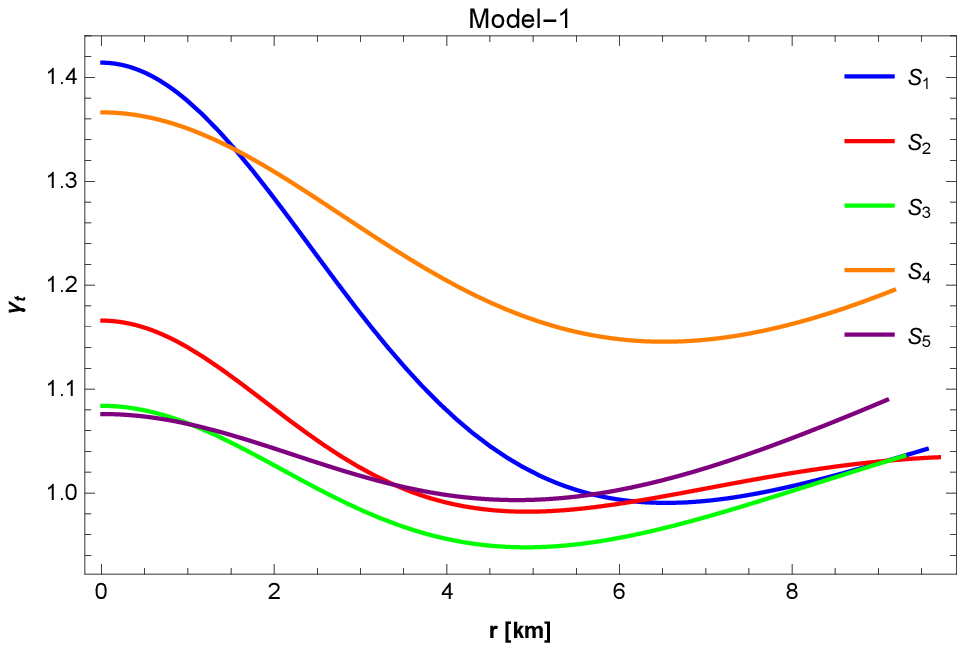,width=0.38\linewidth} &
\end{tabular}
\caption{{Evolution of  $\gamma_{r}$ and $\gamma_{t}$ with respect to ``r" for Model-1}.}
\label{Fig:15}
\end{figure}

\begin{figure}[h!]
\begin{tabular}{cccc}
\epsfig{file=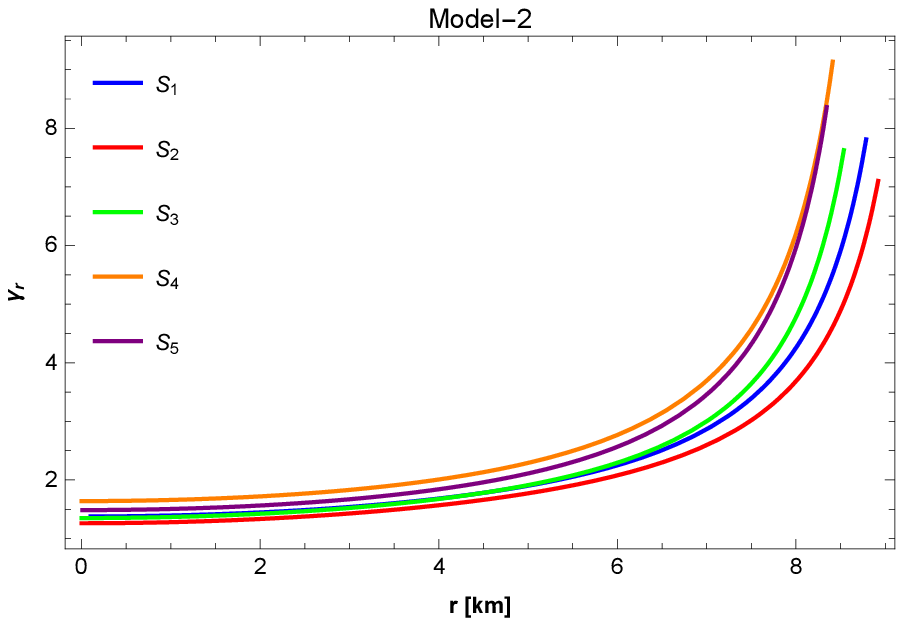,width=0.38\linewidth} &
\epsfig{file=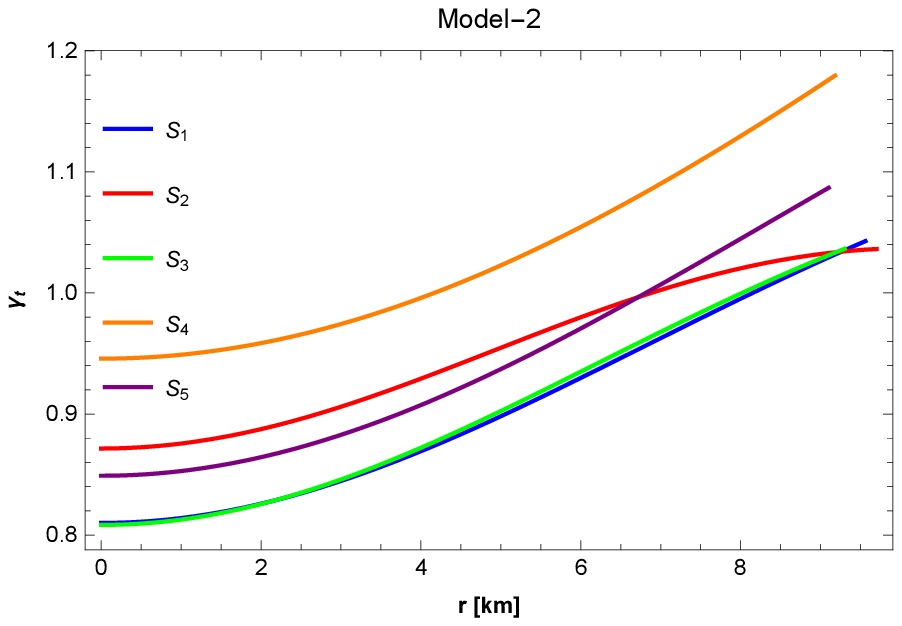,width=0.38\linewidth} &
\end{tabular}
\caption{{Evolution of  $\gamma_{r}$ and $\gamma_{t}$ with respect to ``r" for Model-2}.}
\label{Fig:16}
\end{figure}

Adiabatic index studied to describe the stiffness of EoS, for a given energy density. It provides verification about stability of both relativistic and non relativistic stellar objects. Chandrasekhar \cite{Chand} developed the idea of the dynamical stability against infinitesimal radial adiabatic perturbation of the stellar system. His concept has been successfully tested for isotropic as well as anisotropic stellar spheres, by many authors \cite{HeinHil,Hill,Hor,Don,Silva,Bomb}. For the stability of the compact stars models the value of adiabatic index must be greater than $4/3$. The expressions of the adiabatic index corresponding to radial and transverse pressure components are defined as
\begin{equation}
\gamma_{r}=\frac{\rho+p_{r}}{p_{r}}(\frac{dp_{r}}{d\rho})=\frac{\rho+p_{r}}{p_{r}}{v_{r}}^{2},~~~~~~~\gamma_{t}=\frac{\rho+p_{t}}{p_{t}}(\frac{dp_{t}}{d\rho})=\frac{\rho+p_{t}}{p_{t}}{v_{t}}^{2}.
  \end{equation}
Fig. $\ref{Fig:15}$ and Fig. $\ref{Fig:16},$ shows that this stability condition satisfied everywhere for our proposed two models.

\section{Conclusion}
In order to explore a new family of embedded class-I solutions in the anisotropy background, we assume two realistic $f(R)$ gravity models by using the Karmarkar condition. Since the Karmarkar condition reduced the solution-generating method of EFE to a single metric potential by providing a connection between $g_{rr}$ and $g_{tt}.$ For this purpose, we assume a metric potentials $e^{\lambda}=1+ \frac{a^{2}r^{2}}{(1+br^{2})^{4}}$ and by employing Karmarkar condition we obtained second metric potential given as $e^{\nu}={\Big[A-\frac{aB}{2b(1+br^{2})}\Big]}^{2}$, where $a$, $b$, $A$ and $B$ are arbitrary constants. Further, we calculate the unknown constants by using matching conditions between the interior and Schwarzschild exterior geometries. To verify that our obtained solution are physically acceptable, we analyze the physical response of five individual compact stars.\\
The main goal of our investigation is to generate new solutions of $f(R)$ gravity models in the frame of anisotropic matter source. The significant outcomes are enlisted below.
\begin{itemize}
    \item Fig. $\ref{Fig:1},$ presents that the graphical behavior of both the metric potential $g_{rr}=e^{\lambda}$ and $g_{tt}=e^{\nu}$ are finite, positive, singularity free and fulfil the conditions, i.e., $e^{\lambda(r=0)}=1$  and $e^{\nu(r=0)}={\Big[A-\frac{aB}{2b}\Big]}^{2}$. It has been observed that both metric potentials are monotonically increased and attain maximum values at boundary, which provides evidence that considered models indicate satisfaction outcomes.
    \item The variation of $\rho,$ $p_{r}$ and $p_{t}$ regarding ``r" for both the models are finite and regular at the center. One can easily notice from Fig. $\ref{Fig:2}$ and Fig. $\ref{Fig:3}$ that the corresponding plots attain highest value at the center and express decreasing response toward boundary, which confirm that our proposed models are stable.
    \item The gradient of $\rho,$ $p_{r}$ and $p_{t}$ has been shown in Fig. $\ref{Fig:4}$ and Fig. $\ref{Fig:5},$ which exhibit the consistent nature of our proposed models.
    \item From Fig. $\ref{Fig:6},$ we can easily verify that anisotropy remains positive throughout the compact stars. This behavior  shows that the nature of anisotropy is repulsive which confirms the existence for the compact objects.
    \item From illustration of energy conditions presented in Fig. $\ref{Fig:7}$ and Fig. $\ref{Fig:8},$ we noted that all energy bonds are satisfied.
    \item Fig. $\ref{Fig:9}$, shows the balancing behavior of ($\mathcal{F}_{g}$), ($\mathcal{F}_{h}$) and ($\mathcal{F}_{a}$) of our physical acceptable structure.
    \item For consistent nature of the stellar objects, the graphical response of EoS should honour the constraints $0<w_{r},~w_{t}<1.$ The corresponding graphical illustration provided in Fig. $\ref{Fig:10}$ and Fig. $\ref{Fig:11},$ revealed the stability of our system.
    \item For compact stars, the variation of velocity of sounds $v^2_{r}$ (for radial component) and $v^2_{t}$ (for transverse component) plots should lies within $[0,1].$ From Fig. $\ref{Fig:12}$ and Fig. $\ref{Fig:13}$, we observed that our considered model obeys the causality conditions. In addition, the Herrera cracking concept i.e. $0\leq |v^2_{r}-v^2_{t}|\leq1$, for both the models are consistent, which verify the stability of our system.
     \item It has been observed from Fig. $\ref{Fig:14},$ that the graphs of mass function and compactness parameter obey monotonically increasing responses, while on the other hand, the graphical illustrating of redshift analysis revealed monotonically decreasing behavior, this confirms that our chosen models execute the stability criteria.
      \item  The adiabatic index values of $\gamma_{r}$ and $\gamma_{t}$ for both models are greater than $4/3$, shown in Fig. $\ref{Fig:15}$ and Fig. $\ref{Fig:16},$ which reconfirms the stability of our system.
\end{itemize}
As a concluding remark, in this paper we have favorably presented the well stable and singularity free stellar system, which is appropriate to exhibit the anisotropic nature of the compact stars, by using the Karmarkar condition. Moreover, it is worthwhile to mention here that our presented outcomes are equivalent to the results which were obtained by Bhar \cite{Bhar4} in the context of $GR$.

\end{document}